\documentclass[]{aa} 
\usepackage{graphicx}
\begin{document}
   \title{Molecular Gas in NUclei of GAlaxies (NUGA)}

   \subtitle{I. The counter-rotating LINER NGC\,4826\thanks{Based on 
observations carried out with the IRAM Plateau de Bure Interferometer. 
IRAM is supported by INSU/CNRS (France), MPG (Germany) and IGN (Spain)}}

   \author{S. Garc\'{\i}a-Burillo
          \inst{1}
          \and
	   F. Combes
          \inst{2}
          \and
	   L. K. Hunt
          \inst{3}
          \and
	   F. Boone
          \inst{4}
          \and
	   A. J. Baker
          \inst{5}
          \and
	   L. J. Tacconi
          \inst{5}
          \and
	   A. Eckart
          \inst{6}
          \and
           R. Neri
          \inst{7}
          \and
	   S. Leon
          \inst{8}
          \and
	   E. Schinnerer
          \inst{9}
          \and
           P. Englmaier 
	  \inst{10}
	   }

   \offprints{S.Garc\'{\i}a-Burillo}

   \institute{Observatorio Astron\'omico Nacional (OAN)-Observatorio de Madrid,
Alfonso XII, 3,
28014-Madrid, Spain \\
              \email{burillo@oan.es}
         \and
             Observatoire de Paris, LERMA, 61 Av. de l'Observatoire, 75014-Paris,
France \\
             \email{francoise.combes@obspm.fr}
	\and
	     Istituto di Radioastronomia/CNR, Sez. Firenze, 
		Largo Enrico Fermi, 5, 50125-Firenze, Italy \\
             \email{hunt@arcetri.astro.it}
	\and
	     Astronomisches Institut der Ruhr-Universit\"at Bochum,
Universit\"atstrasse 150,
D-44780,
Germany.\\
             \email{fboone@astro.ruhr-uni-bochum.de}
	\and
	     Max-Planck-Institut f\"ur extraterrestrische Physik, Postfach 1312,
D-85741 
Garching, Germany\\
             \email{ajb@mpe.mpg.de, linda@mpe.mpg.de}
	\and
	     Universit\"at zu K\"oln, I. Physikalisches Institut, Z\"ulpicherstrasse
77, 
50937-K\"oln, Germany\\
             \email{eckart@ph1.uni-koeln.de}
	\and  
 	 Institut de Radio-Astronomie Millim\'etrique (IRAM), 300 Rue de la Piscine, 
38406-St.Mt.D'H\`eres, France\\
             \email{neri@iram.fr}
	\and
	 Instituto de Astrof\'{\i}sica de Andaluc\'{\i}a (CSIC), C/ Bajo de
Hu\'etor, 24,
Apartado
3004, 18080-Granada, Spain\\
             \email{stephane@iaa.es}
	\and
	 National Radio Astronomy Observatory, P.O. Box 0, Socorro, NM87801, USA\\
             \email{eschinne@nrao.edu}
\and
	 Astronomisches Institut, Universit\"at Basel, Venusstrasse 7, CH-4102,
Binningen,
Siwtzerland\\
             \email{ppe@astro.unibas.ch}
%
France \\
%
\\
	}

\date{Received February 7th, 2003; accepted ---, 2003}

   \abstract{We present new high-resolution observations of the nucleus of the counter-rotating
LINER NGC\,4826, made in the $J$=$1-0$ and $J$=$2-1$ lines of $^{12}$CO with the IRAM Plateau de
Bure mm-interferometer(PdBI).The CO maps, which achieve 0.8$\arcsec$(16\,pc) resolution in the 2--1
line, fully resolve an inner molecular gas disk which is truncated at an outer radius of 700\,pc.
The total molecular gas mass (3.1$\times$10$^{8}$M$_{\sun}$) is distributed in a lopsided nuclear
disk of 40\,pc radius, containing 15$\%$ of the total gas mass, and two one-arm spirals, which
develop at different radii in the disk. The distribution and kinematics of molecular gas in the
inner 1\,kpc of NGC\,4826 show the prevalence of different types of $m$=1 perturbations in the
gas. Although dominated by rotation, the gas kinematics are perturbed by streaming motions related
to the $m$=1 instabilities. The non-circular motions associated with the {\it inner} $m$=1
perturbations (lopsided instability and inner one-arm spiral) agree qualitatively with the pattern
expected for a trailing wave developed outside corotation ('fast' wave). In contrast, the streaming
motions in the {\it outer} $m$=1 spiral are better explained by a 'slow' wave. A paradoxical
consequence is that the inner $m$=1 perturbations would not favour AGN feeding. An independent
confirmation that the AGN is not being generously fueled at present is found in the low values of
the gravitational torques exerted by the stellar potential for R$<$530~pc. The distribution of star
formation in the disk of NGC\,4826 is also strongly asymmetrical. The observed asymmetries,
revealed by HST images of the inner disk, follow the scales of the various $m$=1 perturbations
identified in the molecular gas disk. Massive star formation is still vigorous, fed by the
significant molecular gas reservoir at
R$<$700~pc. There is supporting evidence for a recent large mass inflow episode in NGC\,4826. The
onset of $m$=1 instabilities of the type observed in NGC\,4826 may be a consequence of secular
evolution of disks with high gas mass contents. 

These observations have been made in the context of the {\it NUclei of GAlaxies} (NUGA) project,
aimed at the study of the different mechanisms for gas fueling of Active Galactic Nuclei (AGN).

   }

   \maketitle
%

\section{Introduction}

\subsection{Feeding Active Galactic Nuclei: the Nuclei of Galaxies (NUGA)
project} 

Although there is observational evidence that the majority of 
galaxies contain super-massive black holes in their nuclei, the existence of 
nuclear activity is far from universal (Ho et al. 1997). It is commonly accepted that nuclear
activity results from the feeding of a massive black hole by the infall of gas from its host galaxy.
However, there is no consensus on which mechanisms are responsible for removing the angular
momentum from the gas and driving infall down to scales of tens of pc. Moreover, it is unknown
whether these mechanisms are at work only in active galaxies, or, alternatively, if the key
difference between active and quiescent objects is the availability of gas supply to the nucleus.  
On large scales, several dynamical perturbations induced by galaxy 
collisions, mergers and mass accretion (Heckman et al. 1986)
can effectively drive infall of gas to scales of $\sim 1-3\,{\rm kpc}$.
Numerical simulations have also shown that gravity torques from barred potentials are 
efficient at funneling gas into the center of a galaxy. Observational support for the role
of bars in driving gas inflow at these scales has been found (Sakamoto et al. 1999).
However, when an Inner Lindblad Resonance (ILR) is present, the gas accumulates in rings (Combes
1988;
Piner et al. 1995; Buta \& Combes 1996); subsequently there is virtually no gas infall to the
nucleus, as gravity torques are positive inside the ILR.  
   \begin{figure*}[!th]
   \centering
     \includegraphics[width=16cm]{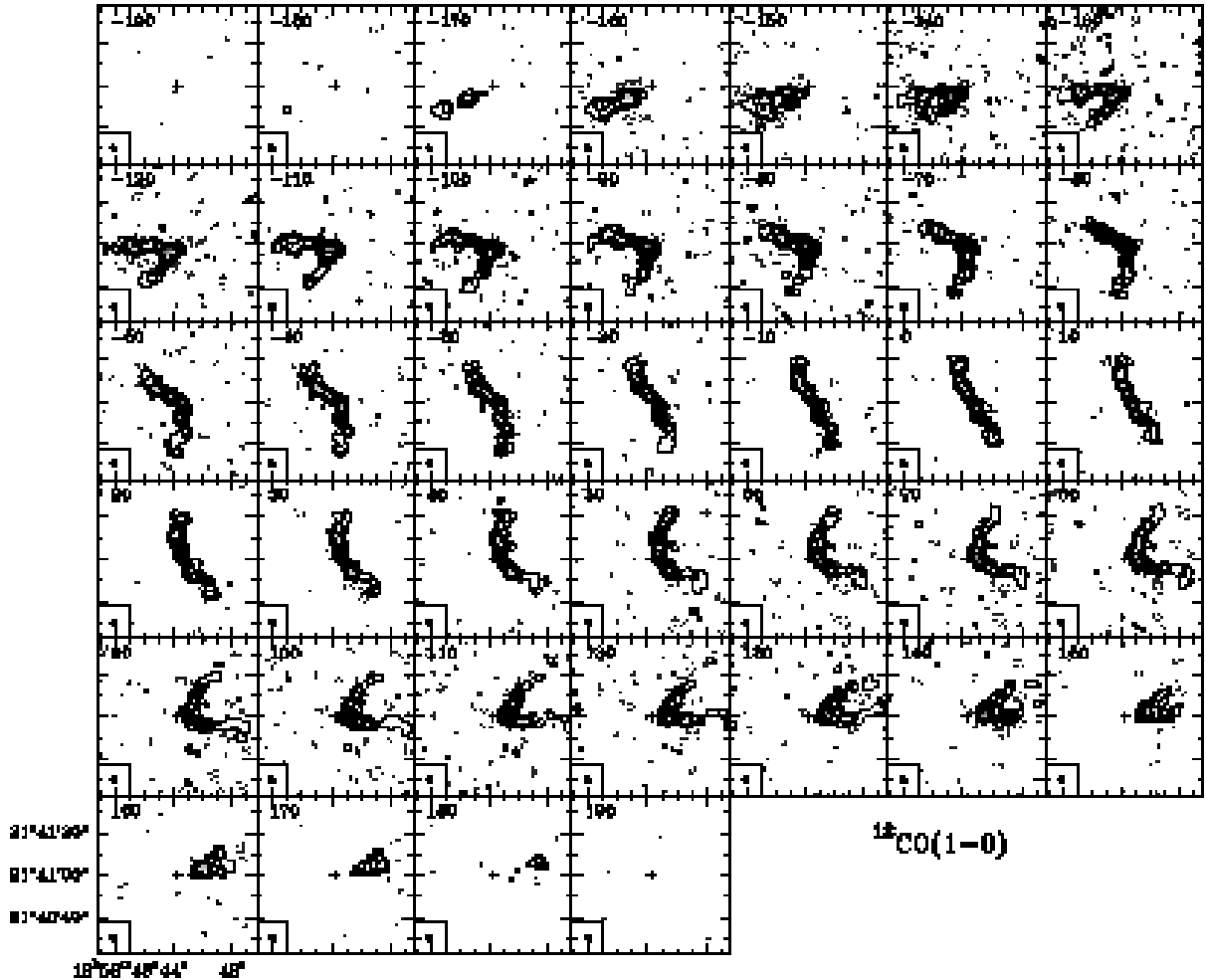}
      \caption{$^{12}$CO(1--0) velocity-channel maps observed with the
PdBI in the nucleus of NGC\,4826 with a spatial resolution of 
3$\arcsec\times$2.4$\arcsec$ at PA=44$^{\circ}$ (beam is plotted as a
filled ellipse in the bottom left corner of each panel). We show a field of view of 75$\arcsec$,
i.e. $\sim$1.8 times the diameter of the primary beam at 115~GHz. The phase tracking center is 
indicated by a cross at $\alpha_{J2000}$=$12^h56^m43.88^s$ 
and $\delta_{J2000}$=$21^{\circ}41'00.1''$. Velocity-channels are displayed from 
v=--190~km~s$^{-1}$ to v=190~km~s$^{-1}$ in steps of 10~km~s$^{-1}$. Velocities are in
LSR scale and referred to v=v$_{o}$=408~km~s$^{-1}$. Contour levels are
--3$\sigma$, 3$\sigma$, 10$\sigma$, 20$\sigma$, 40$\sigma$ and 78$\sigma$ where the 1-sigma
rms $\sigma$=4.0~mJy~beam$^{-1}$. There are very few pixels with negative fluxes below --3$\sigma$
inside the field-of-view.}
         \label{f1}
   \end{figure*}
%
%
  \begin{figure*}[!th]
  \centering
    \includegraphics[width=16cm]{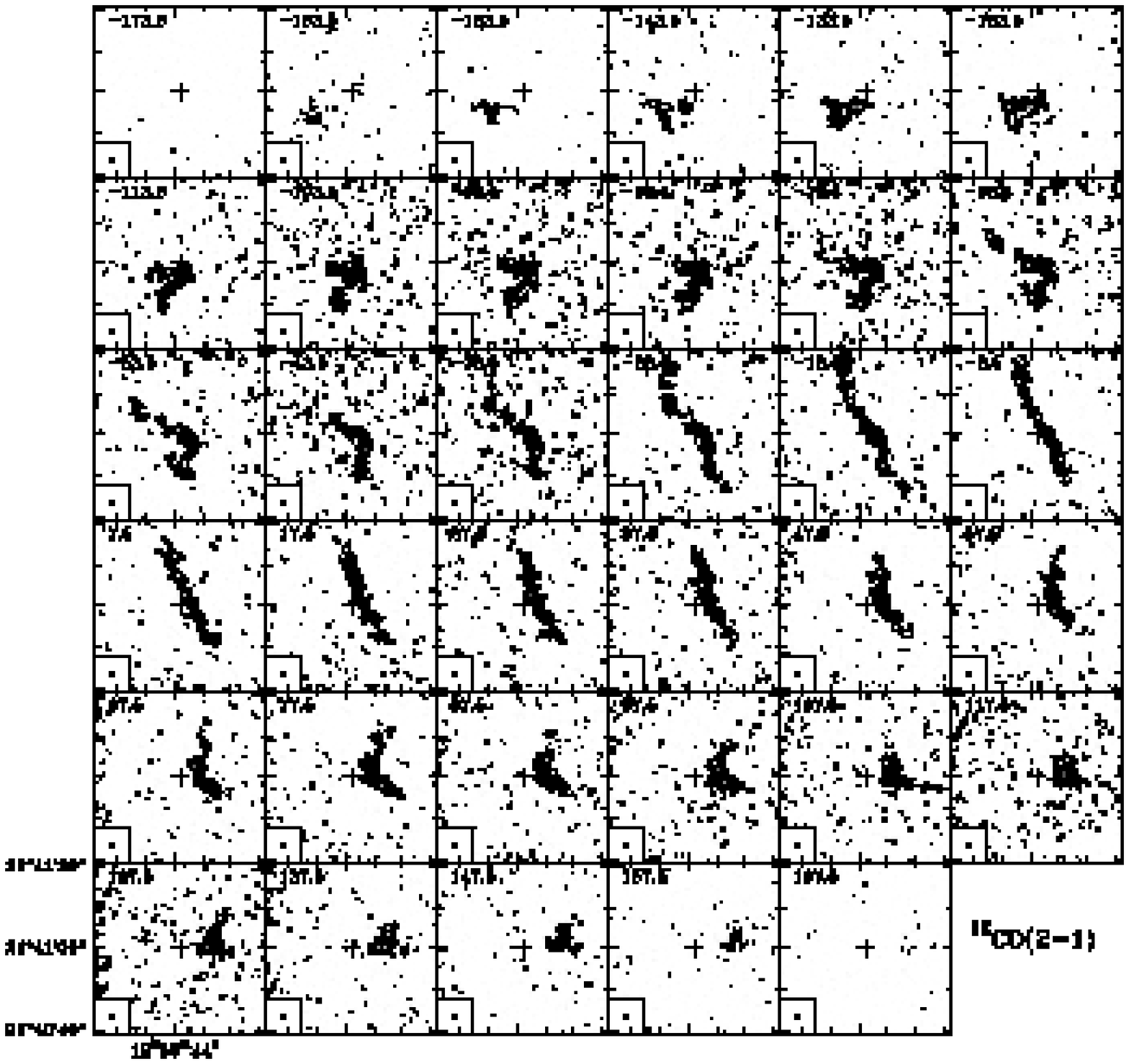}

     \caption{Same as {\bf Fig.~\ref{f1}} but for the 2--1 line of $^{12}$CO. Spatial
resolution reaches 1.1$\arcsec\times$0.8$\arcsec$ at PA=34$^{\circ}$ (beam is plotted as a
filled ellipse in the bottom left corner of each panel). We show a field of view of 
43$\arcsec$, i.e. $\sim$2 times the diameter of the the primary beam at 230~GHz.
Velocity-channels are displayed from v=--172.5~km~s$^{-1}$ to v=167.5~km~s$^{-1}$ in steps of
10~km~s$^{-1}$, with same reference as used in {\bf Fig.~\ref{f1}}. Contour levels are 
--4$\sigma$, 4$\sigma$, 8$\sigma$, 16$\sigma$, 32$\sigma$ and 62$\sigma$, where the 1-sigma
rms $\sigma$=5.7~mJy~beam$^{-1}$. There are very few pixels with negative fluxes below --4$\sigma$
inside the field-of-view.}
         \label{f2}
   \end{figure*}

To explain the mass transfer from the $\sim 1\,{\rm kpc}$ scale 
to radii of tens of pc, we must deal with the non-trivial problem of angular momentum 
removal. Possible solutions have invoked the onset of non-axisymmetric dynamical 
perturbations such as nested nuclear bars (Shlosman et al. 1989; Friedli \& Martinet
1993); lopsidedness or $m$=1 instabilities (Shu et al. 1990; Kormendy \& Bender 1999; 
Garc\'{\i}a-Burillo et al. 2000); warped nuclear disks 
(Pringle 1996; Schinnerer et al. 2000a, 2000b); and nuclear spiral density waves 
(Englmaier \& Shlosman 2000).

The study of interstellar gas in the nuclei of galaxies is a fundamental tool 
for understanding nuclear activity and its relation to circumnuclear star formation.
From the theoretical point of view, there is an increasing body of evidence that purely gaseous
density waves (spirals, bars, warps or lopsided instabilities) may be driving gas infall to the
Active Galactic Nucleus (AGN) (Heller \& Shlosman 1994; Elmegreen et al. 1998; Regan \& Mulchaey
1999). Within the central kiloparsec, most of the gas is in the molecular
phase, which makes CO lines the best tracers of nuclear gas dynamics.  
Up to now, CO surveys of galaxies made with single-dish telescopes were 
hampered by insufficient spatial resolution (Kenney \& Young 1988; 
Heckman et al. 1989; Young et al. 1995; Braine et al. 1993; Casoli et al. 1996; 
Vila-Vilaro et al. 1998). Most CO interferometer surveys of 
nearby spirals (Sakamoto et al. 1999; Regan et al. 2001; Helfer et al. 2003) mapped
CO(1--0) disk emission at low spatial resolution (4$\arcsec$--7$\arcsec$) and moderate 
sensitivity (detectability thresholds $\Sigma_{gas}>$180M$_{\sun}$/pc$^{2}$).  
Furthermore, until very recently (Jogee et al. 2001), the published survey samples
have included very few AGN.

A study of the gas fueling in AGN absolutely requires high
($\sim$0.5$\arcsec$--1$\arcsec$) spatial resolution, given the small ($\sim$50--100\,pc) linear
scales involved. We also need high sensitivity to provide a high dynamic range in the 
synthesized maps. The {\it NUclei of GAlaxies} project--NUGA--(Garc\'{\i}a-Burillo et al.
2003) is a high-resolution, high-sensitivity CO survey of a sample of 12
nearby AGN which spans the sequence of activity types 
(Seyfert\,1, Seyfert\,2 and LINERs). The survey is being carried out with the IRAM Plateau
de Bure mm-interferometer (PdBI) in France, which offers the best combination of sensitivity
and resolution crucial for this project. NUGA aims at reaching spatial resolutions
$<$1$\arcsec$ and will allow the acquisition of CO maps with high dynamic ranges for most of the
targets. Our objective is to determine
the distribution and dynamics of molecular gas in the inner 1\,kpc of the
nuclei with resolutions of $\sim$10--50\,pc, and to study systematically the 
different mechanisms for gas fueling of the AGN. 
 
The NUGA project relies on a multi-wavelength approach: the sample of galaxies observed at PdBI 
has been defined based on the availability of high-quality optical and near-infrared (NIR) images. 
These images have been obtained with both ground-based telescopes and the
Hubble Space Telescope (HST). The optical and NIR counterpart of NUGA will provide
the stellar potentials, the star formation history and the dust distributions for the galaxies
in our sample. We will thus be able to study quantitatively the debated AGN-starburst connection.
NIR maps will be used to compute the stellar potentials. These will be the basis for
self-consistent numerical simulations of the gas dynamics (obtained from CO) in real 
case scenarios (defined by the optical and NIR images). The long term aim is to complete a
super-sample of 25-30 objects observed by consortium members, within and outside NUGA, with the IRAM
array. The initial papers will exploit on a case-by-case basis the data obtained in the context of
the NUGA project. 
   \begin{figure}[!th]
   \centering
   \includegraphics[width=8.8cm]{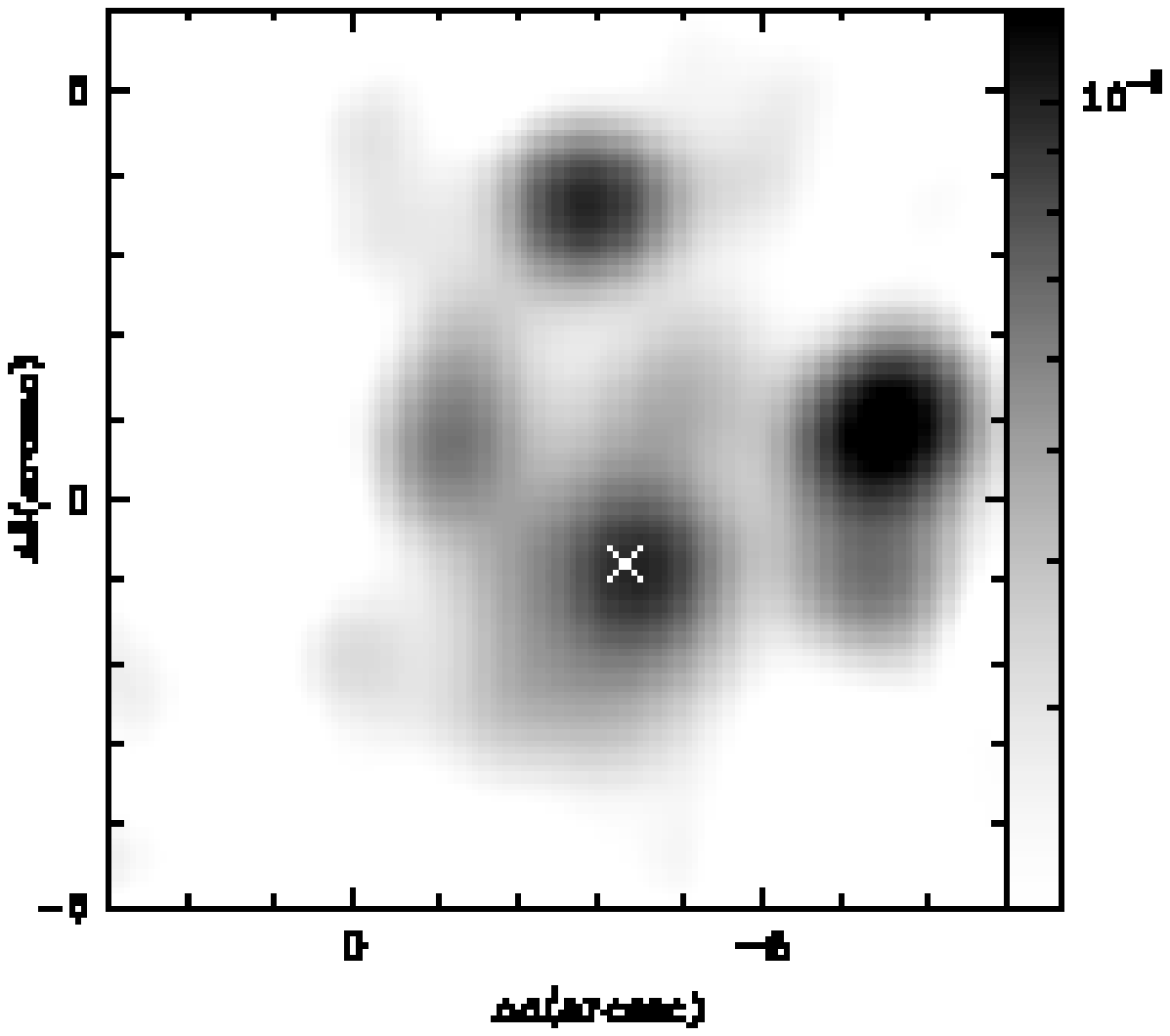}
      \caption{The VLA radio-continuum map at 6\,cm from Turner and Ho (1994), in grey logarithmic
scale from 0.2 to 12mJy~beam$^{-1}$. We highlight the position of the dynamical center derived from
CO (star marker) which coincides with a strong non-thermal source.}
         \label{figure_agn}
   \end{figure}

   \begin{figure}[!h]
   \centering
     \includegraphics[width=8.8cm]{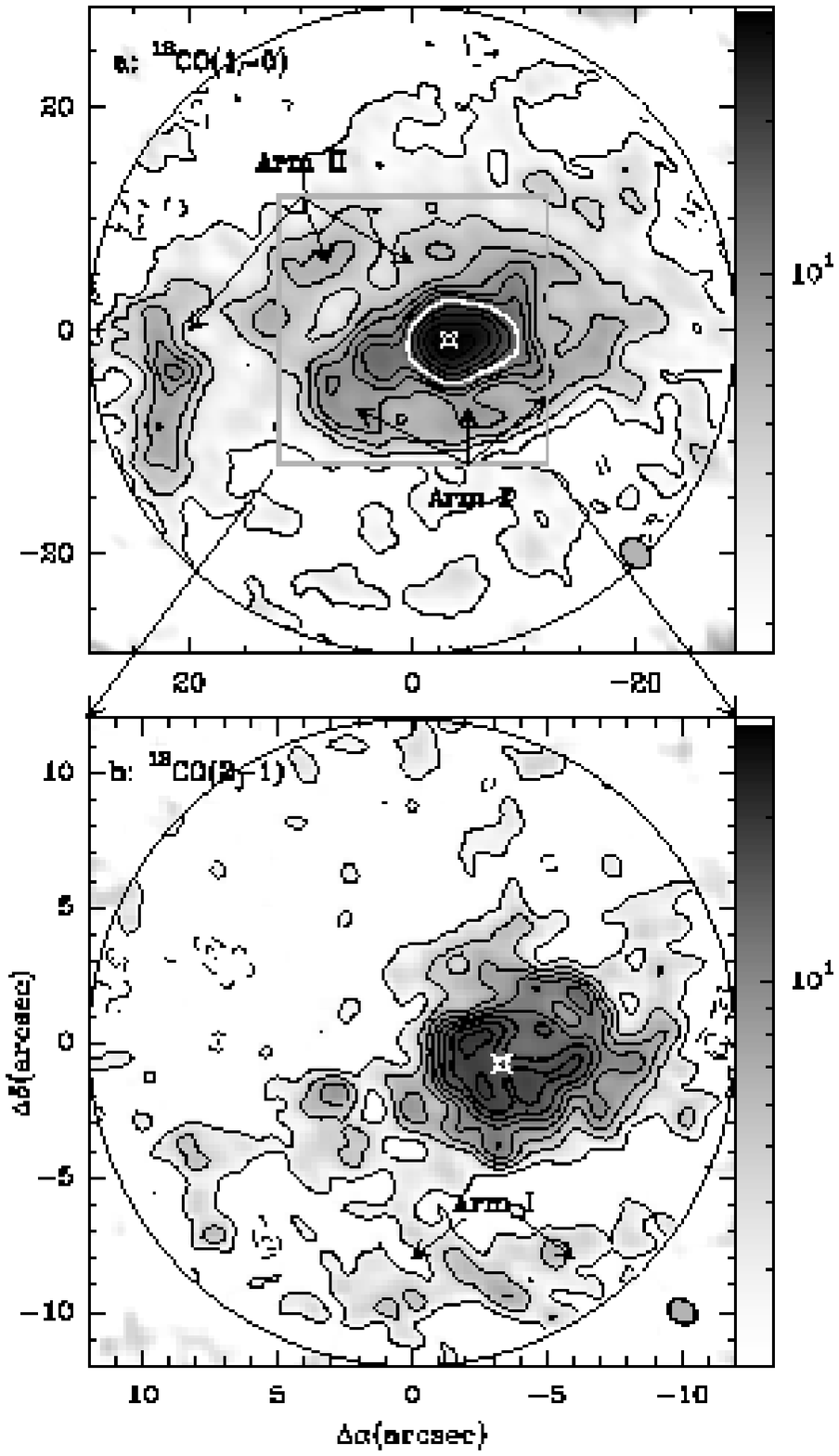}

      \caption{{\bf a)}$^{12}$CO(1--0) integrated intensity contours
in the inner $\sim$58$\arcsec$ of NGC\,4826. Contours beyond the displayed field-of-view have been
screened as S/N ratio is lower than 3 for r$>$29$\arcsec$. $\Delta \alpha$ and $\Delta \delta$ are
offsets (in arcsec) with respect to the phase tracking center. Contours are -1.75, -0.88, 1.75, 4,
6, 8, 10, 12, 15, 18, 22 to 34 in steps of 5Jy\,km\,s$^{-1}$~beam$^{-1}$. The dynamical center is
indicated by the star marker. The white contour highlights the nuclear region here defined as CND.
1-$\sigma$ noise\,=\,0.25Jy\,km\,s$^{-1}$~beam$^{-1}$ at the center of the image (1-$\sigma$
noise\,=\,0.90Jy\,km\,s$^{-1}$~beam$^{-1}$ at the edge of the displayed field-of-view)
{\bf b)} same as {\bf a)} but for the $^{12}$CO(2--1) line. The zoomed view 
shows emission coming from the inner 24$\arcsec$. Contours beyond the displayed field-of-view have
been screened as S/N ratio is lower than 3 for r$>$12$\arcsec$. Contours are 
-2.4 -1.4 2.4 to 19.4 in steps of 2.4Jy\,km\,s$^{-1}$~beam$^{-1}$. Thick contours inside the CND
highlight the asymmetrical pattern. 1-$\sigma$noise\,=\,0.36Jy\,km\,s$^{-1}$~beam$^{-1}$ (1-$\sigma$
noise\,=\,0.8Jy\,km\,s$^{-1}$~beam$^{-1}$ at the edge of the displayed field-of-view). Positions
of Arm I and II are indicated. Beam-sizes are represented by filled ellipses. }

         \label{f3}
   \end{figure}
%

\subsection{The counter-rotator LINER NGC\,4826}

In this first paper, we study the distribution and dynamics of molecular gas
in the nucleus of the LINER NGC\,4826, using high-resolution 
(0.8$\arcsec$--3$\arcsec$) observations made in the 1--0 and 2--1 lines  
of $^{12}$CO with the IRAM array. NGC\,4826, also known as the 'Black Eye' or 
'Evil Eye galaxy' due to its optical appearance, hosts two nested counter-rotating gas disks of
comparable mass. The inner disk extends to a radius of $\sim$50$\arcsec$ and contains 
$\sim$10$^{7}$M$_{\sun}$ of atomic (HI) gas (Braun et al. 1992, 1994) and  
$\sim$2.3$\times$10$^{8}$M$_{\sun}$ of molecular (H$_{2}$) gas (Casoli \& Gerin 1993). The outer
disk, extending from $\sim$80$\arcsec$ to $\sim$9.8$\arcmin$, rotates in the opposite sense to the 
inner gas and contains $\sim$10$^{8}$M$_{\sun}$ of HI (Braun et al. 1992, 1994). 
Rix et al. (1995), by studying the stellar kinematics along the 
principal axes of NGC\,4826, found that the stars rotate at all radii with the
same sense as the inner disk providing strong evidence that stars and gas are coplanar. 
However, the kinematics of the ionized gas, analyzed by Rubin (1994) and
Rix et al. (1995), are considerably more complex. Within radii $<$30$\arcsec$, ionized
gas co-rotates with HI and the stars, followed by a strong  
kinematic disturbance in the region 30$\arcsec<$r$<$100$\arcsec$, where rotation
velocity is close to zero. Most remarkably, Rix et al. (1995) found evidence for radial 
inflow along the minor axis. 

Casoli \& Gerin (1993) made low-resolution $^{12}$CO(1--0) (22$\arcsec$) and
$^{12}$CO(2--1) (12$\arcsec$) maps of NGC\,4826's disk using the IRAM 30m telescope. Their maps
showed that the co-rotating molecular disk ends quite abruptly at a radius of 45-50$\arcsec$,
i.e., at a significantly small extent relative to the optical galaxy size
($D_{25}$=10$\arcmin$). These low-resolution maps could not reveal the small-scale structure and 
kinematics of the compact molecular disk. NGC\,4826 was also observed as a part of the
interferometer surveys of OVRO (Sakamoto et al. 1999) and BIMA-SONG (Helfer et al 2003). These
interferometer maps confirmed that the molecular gas disk of NGC\,4826 extends out to a radius of
700\,pc. The OVRO map of NGC\, 4826, published by Sakamoto et al. (1999), shows tantalizing
evidence of an asymmetrical distribution of molecular gas in the inner 1~kpc.

The maps presented in this paper fully resolve the molecular gas distribution in
NGC\,4826 out to a radius of r=35$\arcsec$(700\,pc). The high spatial resolution 
(0.8$\arcsec$=16\,pc in the 2--1 line) and high dynamic range 
($\sim$70-100) of these observations allow the study of the complex pattern of
gravitational instabilities at work from 1\,kpc to 10-20\,pc scales.
Particular attention is devoted to exploring the influence that counter-rotating 
instabilities may have in driving gas inflow.  
We also compare in detail the CO maps with other gaseous/stellar tracers in order to
obtain a global picture of the gas response to the stellar potential in NGC\,4826.
Finally, we study the distribution of star formation in the nucleus, 
using HST archive broad-band and narrow-band images.

\section{Observations}
\subsection{CO observations \label{obsco}}

Observations of the nuclear region of NGC\,4826 were carried out with the 
IRAM interferometer between December 2000 and January 2002, using the BCD set of 
configurations of the array (Guilloteau et al. 1992). 
We observed simultaneously the J=1--0 and J=2--1 lines of 
$^{12}$CO in a single field centered at $\alpha_{J2000}$=$12^h56^m43.88^s$ 
and $\delta_{J2000}$=$21^{\circ}41'00.1''$; the primary beam size is 
42$''$ (21$''$) in the 1--0 (2--1) line. The spectral correlator was 
split in two halves centered at 115.114\,GHz 
and 230.224\,GHz, respectively, i.e., the transition rest frequencies 
corrected for an assumed recession velocity of v$_{o}(LSR)$=408\,km~s$^{-1}$. 
The correlator configuration covers a bandwidth of 580\,MHz for each line, 
using four 160\,MHz-wide units; this is equivalent 
to 1510\,km~s$^{-1}$(755\,km~s$^{-1}$) at 115\,GHz (230\,GHz). 
The correlator was regularly calibrated by a noise source inserted in
the IF system. Visibilities were obtained using on-source integration 
times of 20 minutes framed by short ($\sim 2$\,min) phase and amplitude
calibrations on the nearby quasars 1308+326 and 3C273. The data were phase 
calibrated in the antenna-based mode. The flux of the primary
calibrators was determined from IRAM measurements and taken as an input
to derive the absolute flux density scale in our map; the latter is estimated to be
accurate to 10\%. The bandpass calibration was carried out using 3C273 and is accurate to
better than 5\%.

The point source sensitivities derived from emission-free channels of 10~km~s$^{-1}$
width are 4.0\,mJy~beam$^{-1}$ in $^{12}$CO(1--0) and 5.7\,mJy~beam$^{-1}$ in
$^{12}$CO(2--1). The image reconstruction was done using
standard IRAM/GAG software (Guilloteau \& Lucas 2000). Unless explicitly stated, we
used natural weighting and no taper to generate the 1--0 line maps with a field of view of
75$\arcsec$ and 0.25$''$ sampling; the corresponding synthesized beam is $3.0''\times 2.3''$,
PA=44$^{\circ}$. We used uniform weighting to generate 2--1 maps with a field of view of
43$\arcsec$ and 0.20$''$ sampling; this enables us to achieve a spatial resolution $<$1$''$ 
($1.1''\times 0.8''$, PA=34$^{\circ}$). 
No 3mm (1mm) continuum was detected towards NGC\,4826,
down to an rms noise level of 0.38~mJy~beam$^{-1}$ (0.67~mJy~beam$^{-1}$) in a 
564\,MHz-wide band centered on 113.6\,GHz (231.7\,GHz). The conversion factors
between Jy\,beam$^{-1}$ and K are 13\,K~Jy$^{-1}$~beam at 115\,GHz, and 25\,K~Jy$^{-1}$~beam at
230\,GHz. By default, all velocities are referred to v$_{o}$ and 
($\Delta\alpha$, $\Delta\delta$) offsets are relative to the phase tracking center. 
Except for channel maps, all displayed maps are corrected for primary beam
attenuation. We will assume a distance to NGC\,4826 of D=4.1$\times$(75\,H$_{o}^{-1}$)\,Mpc (Tully
1988); H$_{o}$\,=\,75\,km\,s$^{-1}$\,Mpc$^{-1}$ implies 1$\arcsec$=20\,pc.
We will assume that the inclination angle of NGC\,4826's disk is $i$=60$^{\circ}$
(Rubin 1994); we derive here a position angle of $PA$=112$\pm$8$^{\circ}$, in rough 
agreement with previous determinations (e.g., PA=120$\pm$5$^{\circ}$; Rix et al. 1995).

\subsection{Optical and near-infrared observations \label{obshst}}

We acquired from the HST archive three broad-band images of NGC\,4826, 
including WFPC2 (F450W ($\sim\,B$ band) and F814W ($\sim\,I$ band)) and
NICMOS (NIC3: F160W ($\sim\,H$ band)); we also obtained a Pa$\alpha$ image (F187N)
as this galaxy was included in the survey of B\"oker et al. (1999).
The optical images were combined with elimination of cosmic rays ({\it crreject}),
and calibrated as described in Holtzman et al. (1995).
Sky values were assumed to be zero. This was checked to be in general a 
good assumption with related NICMOS data (Hunt \& Malkan 2003 in prep), and gives H-band magnitudes
that agree to within 0.1 mag with ground-based data.
The NICMOS images were re-reduced using the best calibration files
and the van der Marel algorithm was used to remove the
``pedestal'' effect (see B\"oker et al. 1999).
The continuum was subtracted from the Pa$\alpha$ image according to the 
precepts explained in B\"oker et al. (1999), but only for the brightest 
pixel values (we excluded roughly half the pixels in the image). 
It was necessary to adjust the slope of F160W-F187N such that the 
nucleus remained $>\,$0 everywhere (see also Garc\'{\i}a-Burillo et al. 2000).
The fields-of-view are 2.7$\arcmin\times$2.7$\arcmin$ and 51$\arcsec\times$51$\arcsec$,
respectively, for the WFPC2 and NICMOS images.

Optical and near-infrared surface-brightness profiles of NGC\,4826
were measured along the minor axis (PA\,=\,22$^\circ$) with a 3$\arcsec$ width, 
and used to derive color profiles after rebinning to a 0.1\,$\arcsec$ pixel scale.
These will be discussed in Sect. \ref{starform}.

\section{Results}

\subsection{Position of the AGN \label{agn}}

Fig.~\ref{f1}-\ref{f2} show the velocity-channel maps of $^{12}$CO(1--0) and $^{12}$CO(2--1)
emission in the central region of NGC\,4826. The typical spider web diagram visible in the
maps is the signature of a spatially resolved rotating disk. The eastern (western) side of
the CO disk is blue (red)-shifted with respect to the reference velocity v$_{o}$. The velocity
field indicates that molecular gas co-rotates with the stars (Rix et al. 1995) and the HI
gas (Braun et al. 1992) to a radius of 700~pc.
We find no evidence of counter-rotating molecular gas in the inner disk. 
The best fits for the center of rotation and systemic velocity are
($\Delta\alpha$,$\Delta\delta$)=(--3.3$''$, --0.8$''$) and
v$_{sys}^{LSR}$=v$_{o}$+8~km~s$^{-1}$=416~km~s$^{-1}$, respectively. Within the errors, the derived
dynamical center coincides with the position where CO lines reach their maximum widths
($\sim$150--200~km~s$^{-1}$). The CO dynamical center lies very close ($\pm$0.2$\arcsec$) to a
strong
non-thermal continuum source present in the 6cm map of Turner \& Ho (1994) (their source number 5;
see also Fig.~\ref{figure_agn} in this work). Furthermore, the dynamical center coincides, within
the errors,
with the position of a X-ray source detected in the high-resolution (0.5$\arcsec$) CHANDRA image of
the nucleus of NGC\,4826 (source S3; PI:Gordon Garmire, id. 411). Therefore, we ascribe the position
$\alpha_{J2000}=12^{\rm h}56^{\rm m}43.64^{\rm s}$, $\delta_{J2000}=21^\circ$40$'$59.3$''$ to both
the dynamical center and the LINER nucleus in NGC\,4826. The systemic velocity derived from CO
[v$_{sys}(LSR)$=416$\pm$5~km~s$^{-1}$=v$_{sys}(HEL)$=408$\pm$5~km~s$^{-1}$]
agrees satisfactorily with the value determined from HI
[v$_{sys}(HEL)$=408$\pm$8~km~s$^{-1}$ from RC3].

   \begin{figure}[!th]
   \centering
   \includegraphics[width=8.2cm]{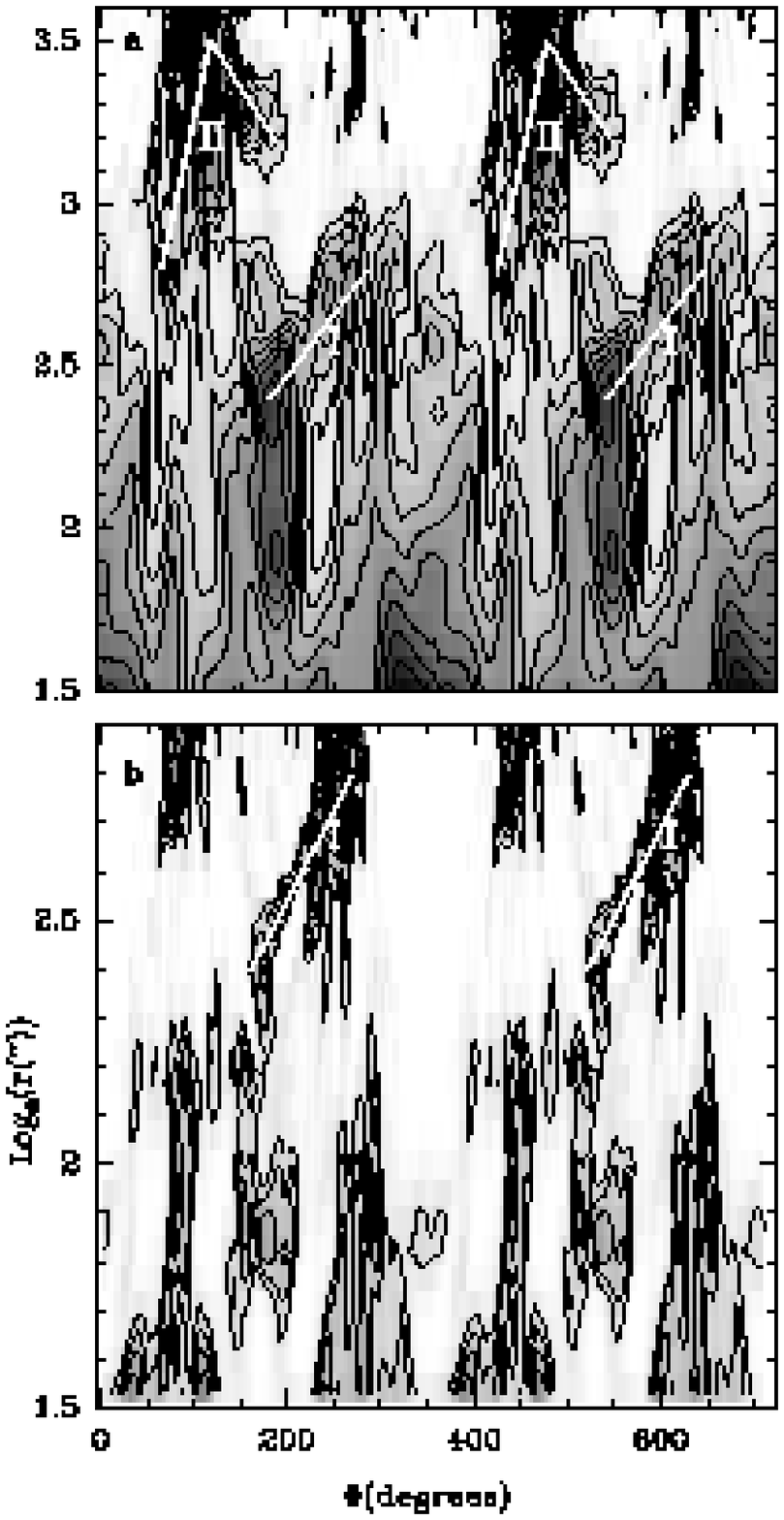}
      \caption{{\bf a)} Peak brightness $^{12}$CO(1--0) map 
deprojected into the galaxy plane in Log$_e(r)$-$\Phi$, 
where r (in arcsec) and $\Phi$ (in degrees) are the polar coordinates. Variable $r$ is the
galaxy deprojected radius in $\arcsec$ relative to the dynamical center and $\Phi$ is the
azimuthal angle in degrees, measured counter-clockwise from the western side of the major axis.  We
assume an inclination of $i$=60$^{\circ}$ and position angle $PA$=112$^{\circ}$ for the NGC\,4826
disk (see Sect. \ref{rot}). The locations of the two $m$=1 spirals (Arms I and II), appearing as two
straight ridges, are high-lighted by the white-shaded lines. {\bf b)} Same as {\bf a)} but for the 
$^{12}$CO(2--1) line. Arm I and the beginning of Arm II are detected.}

         \label{f4}
   \end{figure}

\subsection{The NGC\,4826's molecular gas disk \label{comorph}}

\subsubsection{Molecular gas masses \label{gmass}}

The velocity-integrated $^{12}$CO(1--0) flux within the 42$''$ primary beam field
corrected for primary beam attenuation is S$_{CO}$=1.58$\times$10$^3$Jy~km~s$^{-1}$. Assuming a
CO-to-H$_2$ conversion factor
$X$=N(H$_2$)/I$_{CO}$=2.2$\times$10$^{20}$~cm$^{-2}$~K$^{-1}$~km$^{-1}$~s
(Solomon \& Barrett 1991), the total H$_{2}$ mass derived from the interferometer map 
is M(H$_2$)=2.3$\times$10$^{8}$M$_{\sun}$. 
Including the mass of helium, the total molecular gas mass in the $^{12}$CO(1--0)
Bure field is M$_{gas}$=M(H$_2$+He)=1.36$\times$M(H$_2$)=3.1$\times$10$^{8}$M$_{\sun}$,
in good agreement with the estimate of 3$\times$10$^{8}$M$_{\sun}$ by 
Sakamoto et al. (1999). We estimate that the PdBI map recovers close to 100$\%$ of
the single-dish $^{12}$CO(1--0) flux measured by Casoli \& Gerin (1993) with the IRAM 30m.

The overall distribution of molecular gas is best seen in Fig.~\ref{f3}, which shows the
velocity-integrated intensity $^{12}$CO maps. These were derived by integrating
channels from v$\,=\,-200$ to 200~km~s$^{-1}$. 
Although the $^{12}$CO(1--0) map probably lacks the sensitivity to detect gas
emission beyond r$\sim$35$\arcsec$, we find little evidence that NGC\,4826's molecular disk
extends significantly outside the 42$\arcsec$ primary beam (see Fig.~\ref{f1}).
This confirms the finding of Casoli \& Gerin (1993) (see also Sakamoto et al. 1999) who reported
that molecular gas is confined to the inner 1.6~kpc of NGC\,4826. While our $^{12}$CO(2--1) map
lacks sensitivity outside the central 500~pc of the disk, it gives a sharp image of the molecular
gas distribution in the vicinity of the AGN (Fig.~4b).
 
\subsubsection{The imprint of $m$=1 perturbations in the central kiloparcsec of NGC\,4826} 

Nearly $\sim$15$\%$ of the total molecular gas mass is located in a circumnuclear disk (hereafter
CND) around the AGN (M$_{gas}^{CND}$=3.4$\times$10$^{7}$M$_{\sun}$). This disk has a $\sim$80~pc
average radius and its boundaries are here defined by the eighth intensity contour of
Fig.~4a. The small scale structure of the CND is fully resolved in the 2--1 map (Fig.~4b). At
scales
of $\sim$tens of pc, the CND has a lopsided morphology: there is a pronounced 20--60~pc offset
between the AGN locus and a ridge of $^{12}$CO(2--1) emission which roughly extends from the NE
(--3$\arcsec$, 0.5$\arcsec$) to the W side (--6.5$\arcsec$, 1$\arcsec$) of the CND (see Fig.~4b). A
similar offset between the center of NGC\,4826 and the centroid of dense molecular gas
was identified by Helfer \& Blitz (1997). Their HCN(1--0) map revealed a strong emission peak
at $\sim$(--1.5$\arcsec$,--1.5$\arcsec$), i.e., close to the ridge seen in
$^{12}$CO(2--1), offering supporting evidence that the distribution of molecular gas in the inner
region of the CND (R$<$60~pc) is strongly lopsided. In the outer boundary of the CND
(R$\sim$60--80~pc) two weak
winding spiral arcs can be tentatively identified in the 2--1 map, suggesting that secondary
perturbations are at play. 

Outside the CND, molecular gas emission is also detected along two spiral arms which are
located at noticeably different radii in the disk. This indicates that also at these scales
(100-700~pc) the distribution of molecular gas is asymmetrical. An {\it inner} spiral arm
(hereafter Arm I), identified in both the 1--0 and the 2--1 maps (Fig.~\ref{f3}), extends
south of the CND. The CND and Arm I join at $\sim$(2.5$\arcsec$,--3$\arcsec$). An {\it
outer} spiral arm (hereafter Arm II), visible in the 1--0 map (Fig.~4a), extends north
of the CND, starting from the western side of the disk [$\sim$(--10$\arcsec$, 5$\arcsec$)]. Arm II
ends apparently on the SE side of the disk [at $\sim$(22$\arcsec$,--15$\arcsec$)]. However, CO
emission is detected southward of the CND at declination offsets $\sim$--20$\arcsec$ with a
significant 10$\sigma$ level. Together with Arm II, this S arc would delineate an off-centered
ring.

To highlight the geometry of Arms I and II, we have constructed the $^{12}$CO(1--0)
and $^{12}$CO(2--1) peak brightness intensity maps (sensitive to contrasted 
structures) and deprojected them into the plane of the galaxy. 
Results are displayed in Fig.~\ref{f4}, which represent the deprojected 
brightness maps in polar grid coordinates [Log$_e(r)$,$\Phi$]. Variable $r$ is the
galaxy deprojected radius in $\arcsec$ and $\Phi$ is the azimuthal angle in degrees (see
Fig.~\ref{f4} for details). In this representation, a spiral logarithmic feature would appear as a
straight line with a non-zero slope.
 
   \begin{figure}[!th]
   \centering
   \includegraphics[width=8.8cm]{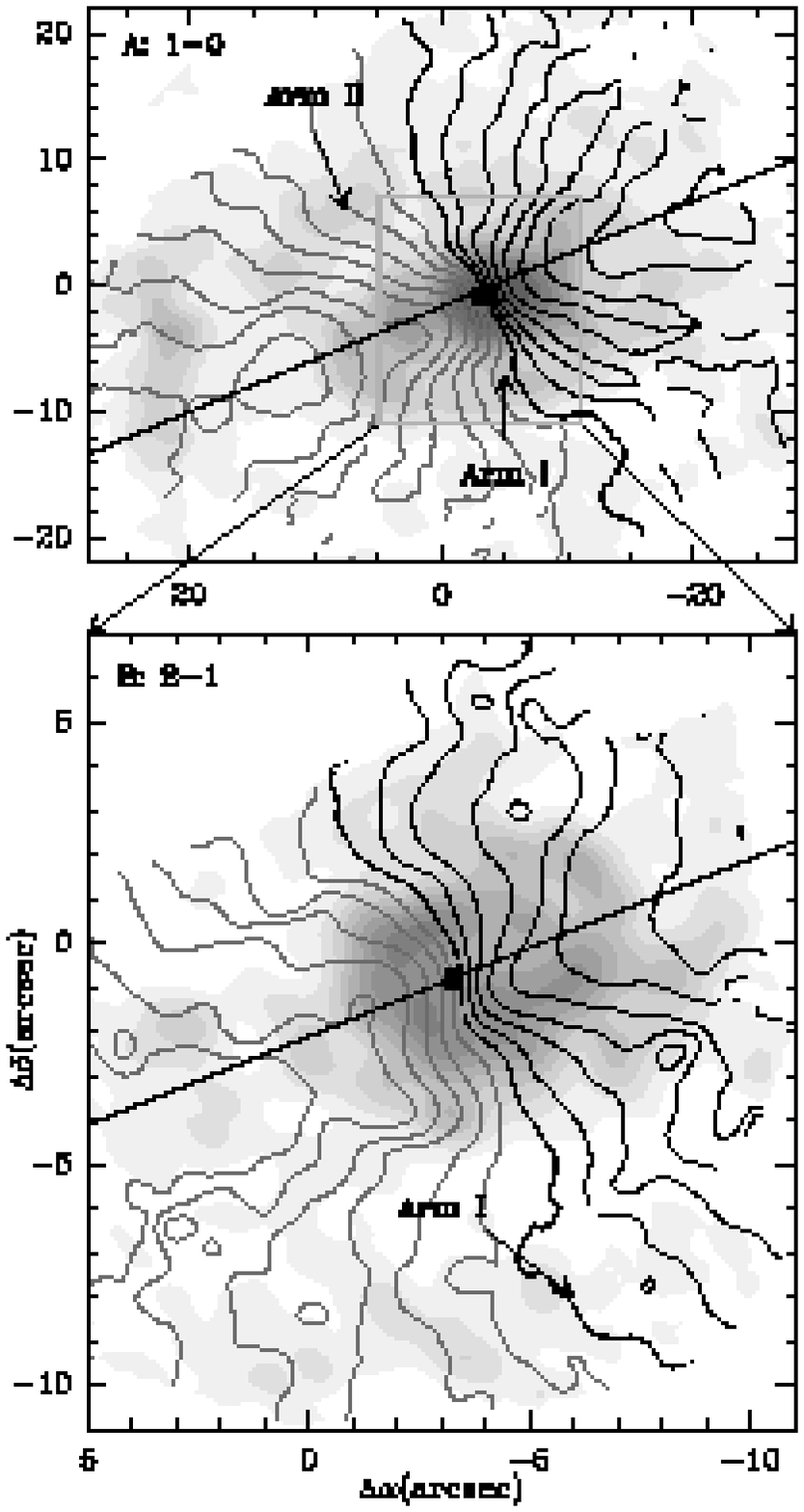}
      \caption{{\bf a)} Overlay of the mean-velocity field derived from the 1--0
data, in line contours, spanning the range (--160~km~s$^{-1}$, 160~km~s$^{-1}$) in steps of
20~km~s$^{-1}$
on the integrated intensity map of $^{12}$CO(1--0). Integrated intensity as in Fig.~\ref{f3}
starting by 1.75~Jy\,km\,s$^{-1}$~beam$^{-1}$. Velocities are here referred to
v$_{sys}(LSR)$=416~km~s$^{-1}$ (thick contour). Solid black (grey) lines are used for positive
(negative) velocities. The straight line at PA=112$^{\circ}$ indicates the position of NGC\,4826's
major axis. The star marks the AGN position. {\bf b)} Same as {\bf a)} but for the $^{12}$CO(2--1)
line. Integrated intensity contours as in Fig.~\ref{f3} starting by
2.4~Jy\,km\,s$^{-1}$~beam$^{-1}$.
Velocities span the range (--120~km~s$^{-1}$, 120~km~s$^{-1}$) in steps of
20~km~s$^{-1}$.}
         \label{f5}
   \end{figure}

   \begin{figure}[!th]
   \centering
   \includegraphics[width=8.8cm]{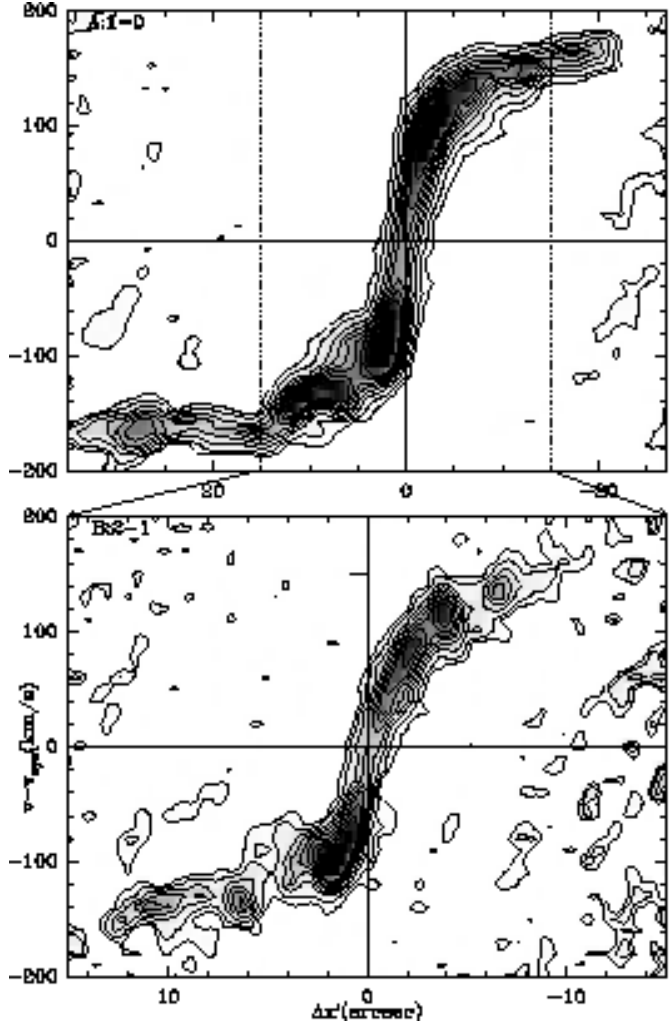}
      \caption{{\bf a)} We show the $^{12}$CO(1--0) position-velocity (p-v)
diagram along the major axis of NGC\,4826 (line contours and grey
scale range: 0.015, 0.035 to 0.27~Jy~beam$^{-1}$ in steps of
 0.035~Jy~beam$^{-1}$). Velocities have been re-scaled to v$_{sys}^{LSR}$=416~km~s$^{-1}$
and x' offsets are relative to the dynamical center.
{\bf b)}: Same as {\bf a)} but for the $^{12}$CO(2--1) line in the inner
r$\sim$15$\arcsec$. Contours are the same as in {\bf a)}.}
         \label{f6}
   \end{figure}

   \begin{figure}[!th]
   \centering
   \includegraphics[width=6.5cm]{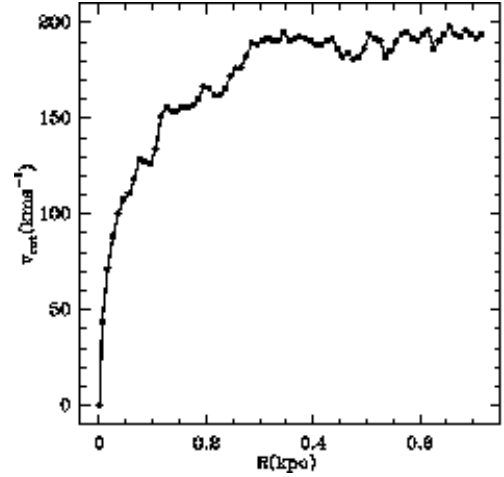}
      \caption{Rotation curve obtained from the combined 1--0 + 2--1 CO data in 
NGC\,4826 to a radius of R=700~pc.}
         \label{f7}
   \end{figure}

   \begin{figure}[!th]
   \centering
   \includegraphics[width=8.8cm]{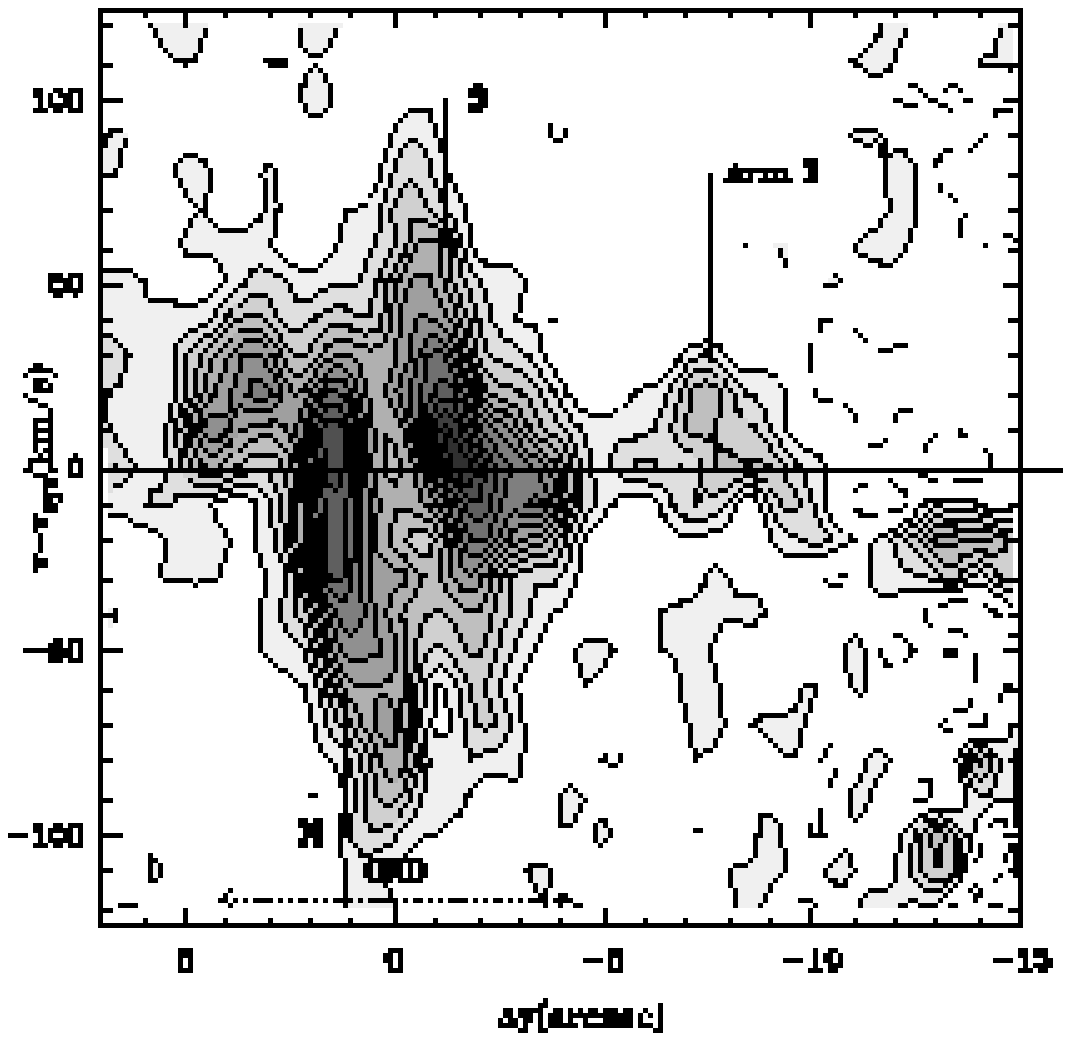}
      \caption{$^{12}$CO(2--1) position-velocity plot taken along the NGC\,4826's minor axis.
We indicate the position of Arm I, and the N/S crossings of the CND lopsided instability. The gas
kinematics show local signatures of outflow motions at these positions.}
         \label{fminor}
   \end{figure}

%

As shown in Fig.~\ref{f4}, Arm I develops between 2.4$<$Log$_e(r)$$<$2.8 
(11$\arcsec$$<$r$<$16$\arcsec$) and 160$^{\circ}$$<$$\Phi$$<$280$^{\circ}$. 
Arm I shows {\it only} a 2$\times\pi$--periodicity along $\Phi$. In contrast to two-arm ($m$=2)
spiral arms, one-arm ($m$=1) spirals are not $\pi$-periodic in azimuth. We can
derive an average pitch angle ($p$) of the spiral from
tan($p$)=--$\Delta$Log$_e(r)$/$\Delta\Phi$ along the arm; $p$ is the angle
between the spiral and a circle at a given radius, measured
counter-clockwise. This gives $p\sim$167$^{\circ}$ for Arm I.
 The dust lane seen in optical images of NGC\,4826 is a projected 
foreground layer situated on the northern side of the galaxy, identifying this as the
nearer side. According to the observed sense of rotation of the gas, we therefore conclude
that Arm I is {\it trailing} with respect to the gas and the stellar flows.
Interestingly, this is in agreement with the morphological analysis by
Walterbos et al. (1994), who also deduced that the inner spiral structure must be
trailing inside R$\sim$2.7~kpc.

Arm II is seen in the outer disk in the range 2.8$<$Log$_e(r)$$<$3.5
(16$\arcsec$$<$r$<$33$\arcsec$)
and 65$^{\circ}$$<$$\Phi$$<$190$^{\circ}$. Arm II  shows a 
low order 2$\times\pi$--periodicity along $\Phi$, characteristic of $m$=1 instabilities. 
A similar analysis as done for Arm I indicates that the 
pitch angle of Arm II changes midway from $p\sim$146$^{\circ}$ (trailing) 
to $p\sim$10$^{\circ}$ (leading). This sudden change of orientation might suggest that at this
location (R$\sim$600~pc) two spiral-like perturbations are meeting in the disk.

   \begin{figure}[!th]
   \centering
   \includegraphics[width=7.5cm]{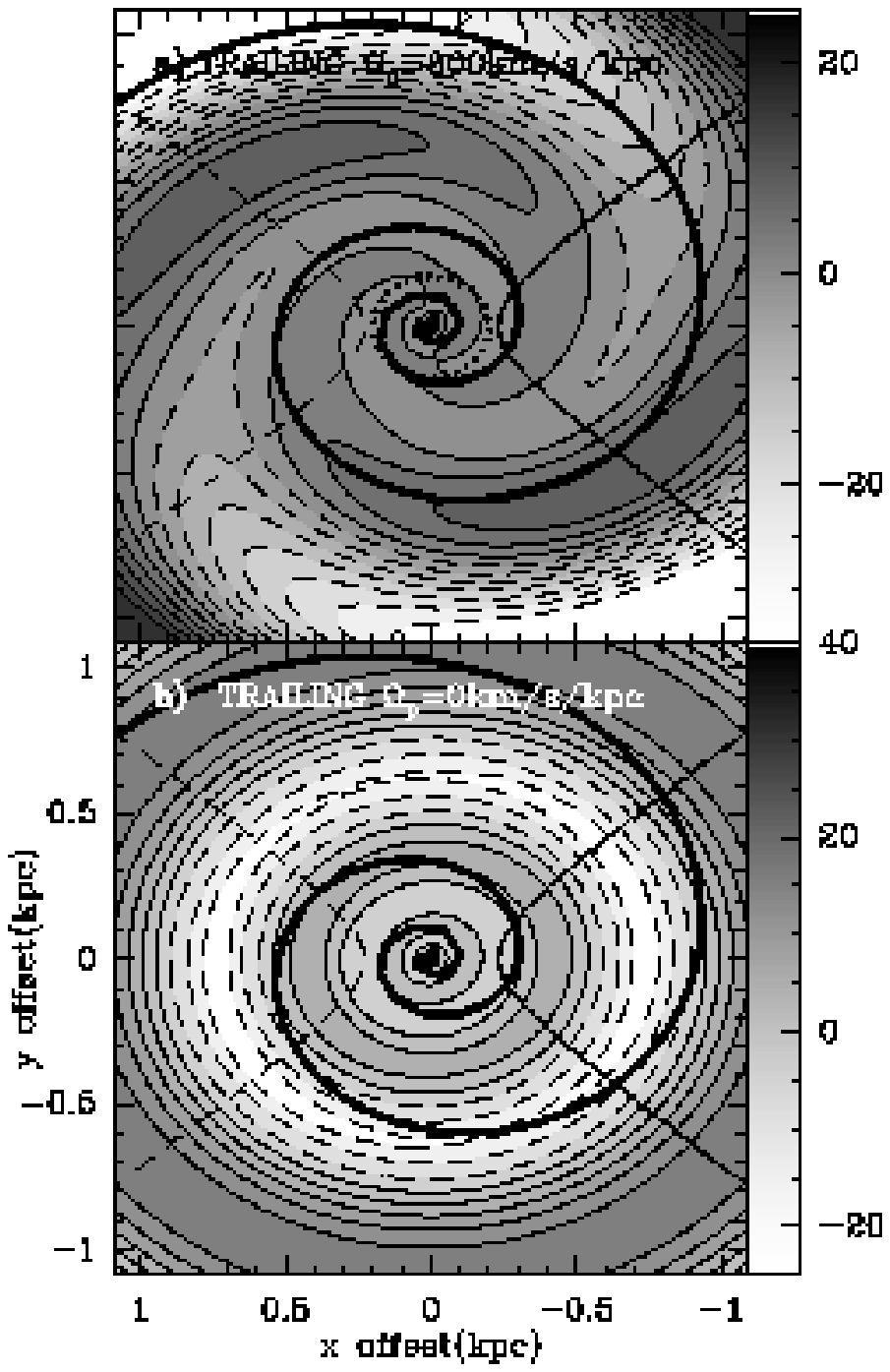}
      \caption{We display the velocity perturbations of the gas flow, projected into the plane of
the sky (v$_{pert}$), due to a {\it fast}
($\Omega_p$=800~km~s$^{-1}$kpc$^{-1}$) {\bf(a)} and a {\it slow} 
($\Omega_p$=0~km~s$^{-1}$kpc$^{-1}$) {\bf(b)} $m$=1 trailing spiral wave. 
 $x$ and $y$ axes are parallel to the
major and minor galaxy axes (x$>$0 eastwards, y$>$0 northwards), and we assume i=--30$^{\circ}$
(i.e., northern side is closer to us). Thin line contours and gray scale range from --10, --8 to 16
in steps of 2~km~s$^{-1}$ (dashed contours for negative values). Isovelocities v=--80~km~s$^{-1}$
(dashed line) and v=80~km~s$^{-1}$ (thick line) define the orientation of circular rotation in the
disk (clockwise). The potential minimum locus is represented by the logarithmic spiral. The
position of corotation is indicated by the dotted circle in the top panel.}
         \label{f8}
   \end{figure}

   \begin{figure}[!th]
   \centering
   \includegraphics[width=8.8cm]{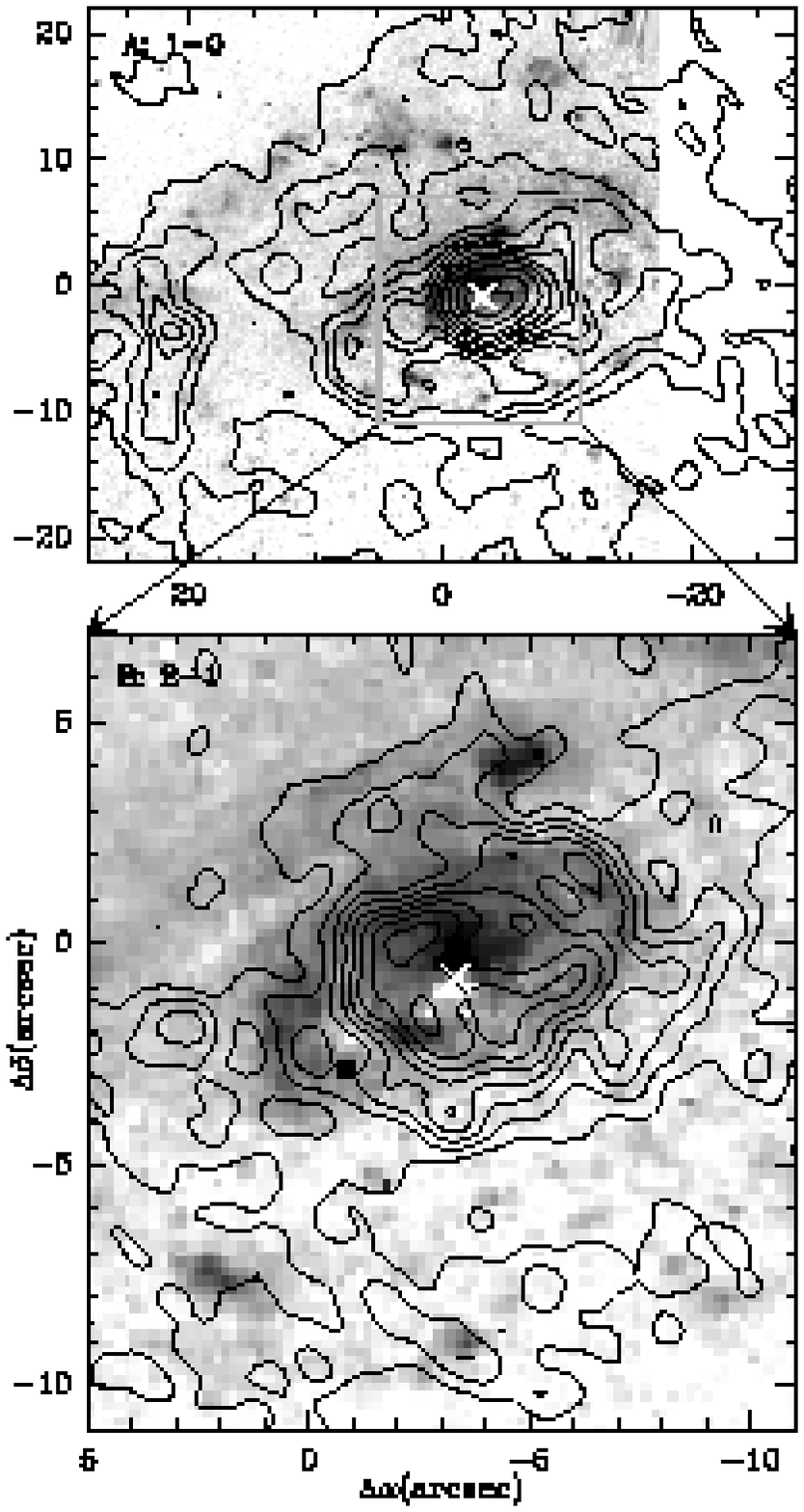}
      \caption{We display the overlay of the Pa${\alpha}$ integrated intensity 
image, derived from the HST F187N and F160W NICMOS filters (grey logarithmic scale
ranges from 3.1\,10$^{-12}$--2.1\,10$^{-11}$erg~s$^{-1}$cm$^{-2}$pix$^{-1}$ in the top panel and
from 7.7\,10$^{-13}$--3.1\,10$^{-11}$erg~s$^{-1}$cm$^{-2}$pix$^{-1}$ in the bottom panel;
grey scale increases from light to dark), and the integrated intensity maps obtained for the 1--0
(top) and 2--1(bottom) lines of $^{12}$CO in the disk of NGC\,4826 (same contours as
Fig.~\ref{f5}).}
      \label{f9}
   \end{figure}
In summary, the distribution of molecular gas in the inner 1~kpc of NGC\,4826 shows the 
prevalence of different types of asymmetrical $m$=1 perturbations in the gas disk. Besides two 
one-arm trailing spirals--Arms I and II--which develop in the outer region (radii
$\sim$100--700~pc), we have detected a lopsided CND. 
This indicates that $m$=1 perturbations extend to radial distances as small as
$\sim$20--60~pc from the AGN.  As shown below, the gas kinematics in the inner 1~kpc reveal
the presence of non-circular motions which are related to the various detected $m$=1
perturbations. A detailed analysis of these is developed in Sect. \ref{modes} to give an insight
into the origin of $m$=1 perturbations in NGC\,4826.

\subsection{Kinematics of NGC\,4826's molecular gas disk \label{kin}}

\subsubsection{The rotation curve and the dynamical mass \label{rot}}

We show in Fig.~\ref{f5} the mean velocity field maps obtained from the $^{12}$CO(1--0) 
and $^{12}$CO(2--1) data in the nucleus of NGC\,4826 (derived with 2-$\sigma$ clipping). 
The kinematics of molecular gas at radii R$<$700~pc are consistent with those of a 
disk in direct rotation with respect to the stars.  

A CO rotation curve (v$_{rot}$) has been derived from 
position-velocity (p--v) diagrams taken along the kinematical major axis of 
NGC\,4826 (Fig.~\ref{f6}). We identify the major axis position angle as
PA=112$\pm$8$^{\circ}$; this value yields a larger line-of-sight velocity gradient
 within R=700~pc than does a p--v cut through the dynamical center at any other angle. 
Our determination roughly agrees with previous findings 
based on H$\alpha$ and stellar kinematics (PA=120$\pm$5$^{\circ}$; Rix et al 1995).
We have calculated the terminal velocities, and from these, the rotation curve, by
fitting gaussian profiles to the spectra across the major axis. The velocity centroids, 
corrected for inclination $i$=60$^{\circ}$, give v$_{rot}$ for each offset 
along the major axis. Rotation curves derived from both sides of the major axis 
do not differ significantly within the errors. Therefore, we have derived v$_{rot}$ as a function of
radius by averaging values from the W and E sides of the major axis. Data for both lines were 
noise-weighted averaged using a radial beaming of $\Delta$r=0.5$\arcsec$. 
 
The resultant v$_{rot}$ (Fig.~\ref{f7}) shows a steep increase to the edge of the CND 
(R$\sim$70--80~pc), where v$_{rot}\sim$120-125~km~s$^{-1}$. This trend is followed by a 
gradual increase up to a radius of R$\sim$350~pc, after which v$_{rot}$
remains flat at $\sim$190~km~s$^{-1}$. The rotation velocity of molecular gas follows the
stellar and the ionized gas motions inside R$\sim$700~pc (see Fig.~3 and 6 of Rix et al 1995).

The total mass inside the CND can be inferred using M(R)=C$\times$R$\times$v$_{rot}^2$/G, 
where G is the constant of gravitation, M(R) is the mass inside a sphere of
radius R, and C is constant varying between 0.6 and 1, depending on the disk mass model
assumed (Lequeux 1983). If we take a value of C=0.8 intermediate between the values
appropriate for spherical (1) and flat disk (0.6) distributions, 
M(R=70~pc)$\sim$2.3$\times$10$^{8}$M$_{\sun}$. 
This implies that the molecular gas mass fraction inside the CND is $\sim$15$\%$,
decreasing to $\sim$5$\%$ within R=700~pc.

With the value estimated for the rotation curve at a radius R$\sim$10\,pc
(v$_{rot}\sim$58~km~s$^{-1}$) we can set an upper limit of $\sim$8$\times$10$^{6}$M$_{\sun}$ for the
mass of the putative super-massive black hole in the nucleus of NGC\,4826.

\subsubsection{Streaming motions and $m$=1 modes \label{modes}}

Although mainly characterized by circular rotation, the gas kinematics are perturbed 
by streaming motions. Isovelocities in Fig.~\ref{f5} display a wavy pattern at the passage 
of both Arms I and II.  Deviations from circular motion can be 
identified in Arm I as a systematic kink in the isovelocities which appear 
redshifted; as expected, the shift is most pronounced near the crossing of the galaxy minor axis
where only radial motions have non-zero projection (see Fig.~\ref{fminor}). We find a gradient
across the minor axis from  redshifted velocities in Arm I to blueshifted velocities in the
adjacent interarm region located at larger radii (see Fig.~6a and Fig.~\ref{fminor}). Deprojected
into the galaxy plane, the radial velocities measured on Arm I indicate local outflow motions. 
In contrast, non-circular motions along Arm II indicate that gas velocities 
are systematically redshifted, a local signature of inflow motions (Fig.~\ref{f5}).  

Departures from pure rotation, related to the lopsided instability, also characterize the CND
kinematics. The average gas velocities derived from 2--1 data seem slightly redshifted
(blueshifted) on the southern (northern) crossing of the lopsided feature along the minor axis
(Fig.~\ref{fminor}). Similarly to Arm I, these motions can be interpreted as a local signature of
outflow. However radial streaming motions are neatly reversed in the outer boundary of the CND
(R$\sim$60--80~pc) where two secondary winding spiral arcs are identified in the 2--1 map (see
Fig.~\ref{fminor}).

   \begin{figure}[!th]
   \centering
   \includegraphics[width=8.8cm]{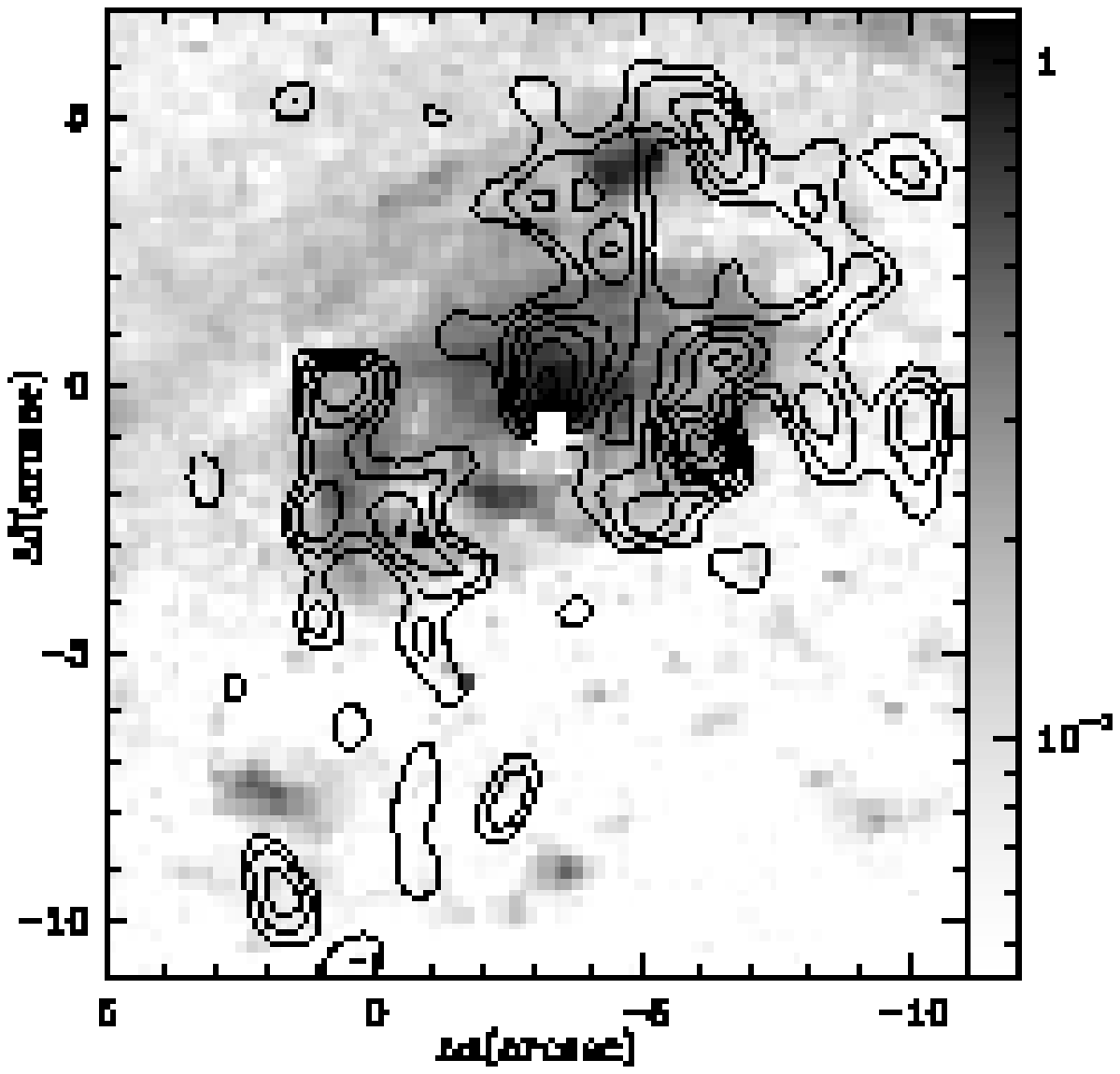}
      \caption{The VLA radio-continuum map at 2cm (mostly thermal emission) 
taken from Turner and Ho (1994) (in contours from 35$\%$ to 95$\%$, in steps of
10$\%$ of the maximum=1~mJy~beam$^{-1}$) is overlaid on the 
Pa${\alpha}$ image of NGC\,4826 (in grey scale, as shown). 
The lopsided star formation in the nucleus has a strong maximum offset by $\sim$16~pc northward from
the position of the AGN (marked with a square).}
         \label{f12}
   \end{figure}

The signatures of the streaming motions expected for trailing/leading waves can be
analyzed in the framework of the linear density wave theory (Shu et al. 1973). 
The sign of velocity perturbations changes close to the minor axis when
crossing from inside to outside the corotation resonance. A comparison with models may give some
insight into the nature of the density waves that may account for the streaming motions observed in
CO. Garc\'{\i}a-Burillo et al. (2000) applied the general formalism to the case of one-arm ($m$=1)
spirals for two extreme values of the pattern speed ($\Omega_p$) illustrating the slow and the fast
mode solutions. In the slow (fast) solution we are mostly inside (outside) corotation of the modes.  
In the case of NGC\,4826, we have explored a fast solution by adopting a pattern speed 
large enough to assure corotation lies well inside the disk of the 'model' galaxy
($\Omega_p$=800~km\,s$^{-1}$~kpc$^{-1}$; this implies R$_{COR}$=200\,pc for a
rotation curve similar to v$_{rot}$). In contrast, a slow mode is here characterized by
$\Omega_p$=0~km\,s$^{-1}$kpc$^{-1}$, i.e., a stationary wave where corotation is pushed outside the
disk. 

We represent in Fig.~\ref{f8} the radial velocity perturbations, projected into the plane of the sky
(v$_{pert}$), for the fast and slow one-arm trailing solutions.  We purposely chose a disk geometry
and spiral parameters qualitatively similar to those of NGC\,4826, apart from a lower inclination
angle adopted here to give a more detailed picture of the velocity field in the model. Moreover, we
can reasonably assume that the gas response (peak gas density) is close to the spiral potential
minima in all cases. A comparison of the two model solutions with the CO observations indicates
that the streaming motions measured on Arm I and, also, on the inner lopsided instability of the
CND match qualitatively the solution of the fast trailing $m$=1 mode (Fig.~\ref{f8}): when the
spiral arm is outside corotation, velocities are redshifted (blueshifted) on the southern
(northern) side of the galaxy minor axis. In contrast, Arm II better matches the solution of the
slow trailing mode (Fig.~\ref{f8}).

A paradoxical consequence is that the inner $m$=1 perturbations, represented by Arm I and the
lopsided CND instability, would not favor gas infall: behaving as fast modes, the gas would gain
angular momentum from the waves and would migrate outwards. In particular the AGN fueling
might be temporarily blocked upon the onset of the inner lopsided instability. The perturbations
identified in the outer boundary of the CND would favor gas inflow, however. This suggests that
inflow and outflow may be globally counterbalanced in the CND. It remains to be determined, however,
if angular momentum removal/gain from these instabilities, either stellar or gaseous, can be
efficient enough to play some role in AGN feeding (see Sect. \ref{grav}).

   \begin{figure*}[!th]
   \centering
   \includegraphics[width=8.5cm]{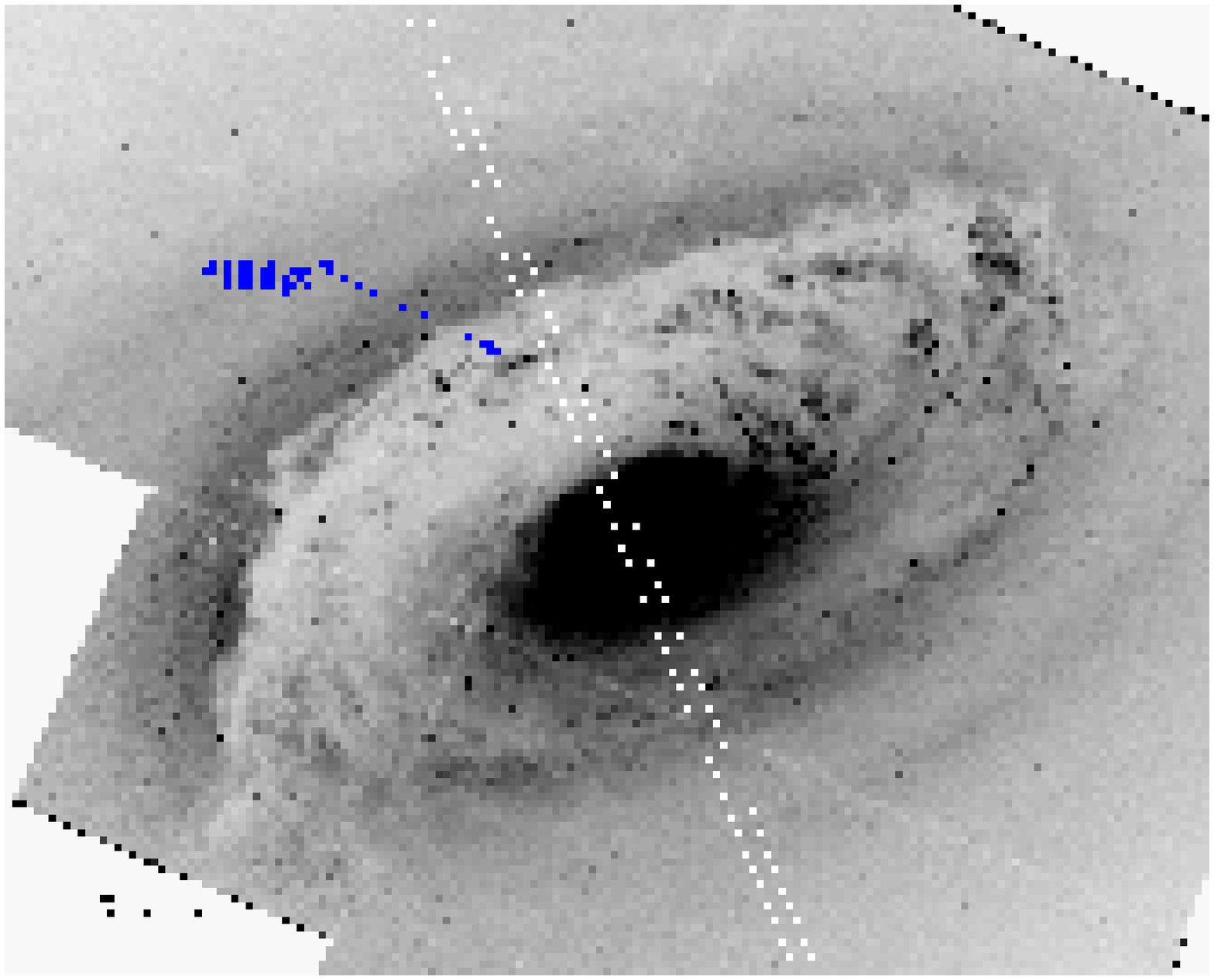}
   \includegraphics[width=8.5cm]{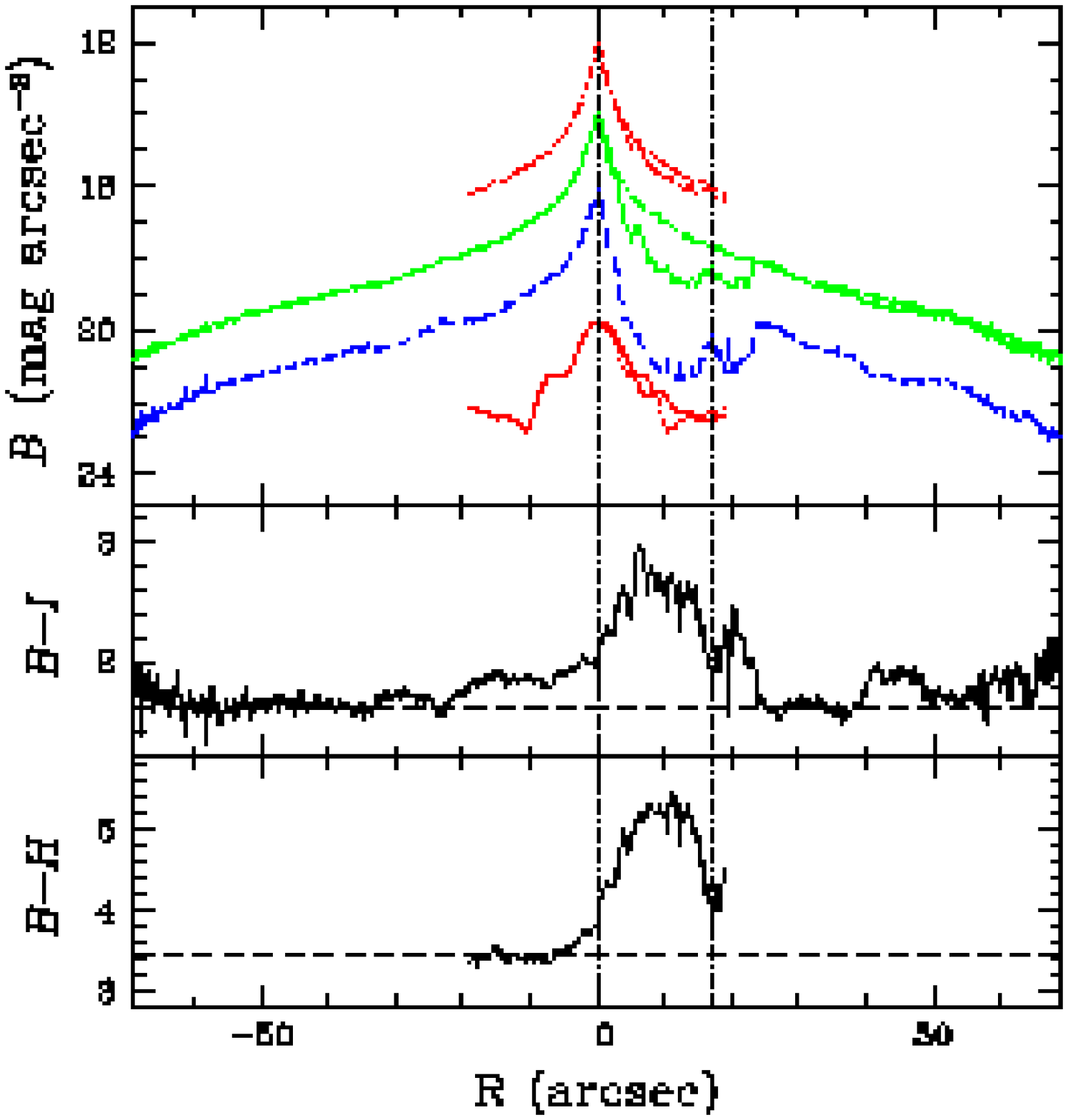}
      \caption{{\bf a)(left)}The F450W image of the galaxy with the ``cut'' aperture superposed. 
The brightness peak is shown with a cross in the black (brightest) central region, while a
bright H{\sc II} region is marked with a cross at a radial distance of 17$\arcsec$. {\bf b)(right)}
The surface brightness profile along the minor axis, together with the F450W--F814W $\sim\,B-I$ and
F450W--F160W $\sim\,B-H$ colors. The center and H{\sc II} region at $r\,\sim\,17\arcsec$ 
shown by crosses in {\bf a)(left)} are denoted here by vertical dotted lines. 
The F160W profile is shown by a dotted line, F814W by a solid line, F450W by a dashed line,
and CO(1-0) by a heavy line (from top to bottom). Y axis
units are magnitudes for all profiles, 'including' CO (in logarithmic scale with an arbitrary
offset). In the top panel (for all curves except that for F450W), we plot the reflection
of the southern half of the surface brightness profile underneath the northern half of
the profile, showing the asymmetry induced by the dust lane. The horizontal dashed lines in the
lower panels show the intrinsic color as estimated from the southern region, i.e., at negative
offsets along the cuts.}
      \label{galaxy}
   \end{figure*}

%

\subsection{CO line ratios \label{ratios}}

To probe the physical conditions of the molecular gas in NGC\,4826, we have studied the variation of
the $^{12}$CO(2--1)/$^{12}$CO(1--0) ratio in areas with significant emission levels in both lines 
(brightness temperatures $>$5$\sigma$). The 2--1 map was first degraded to the spatial resolution of
the 1--0 map within the 21$\arcsec$ primary beam to assure that the two lines sample identical
regions; both maps were also corrected for primary beam attenuation. 

The derived $^{12}$CO(2--1)/$^{12}$CO(1--0) ratio is roughly constant over most of
the disk, with a mean value 0.65$\pm$0.05. This ratio is close to the canonical value observed in 
other spiral disks and is typical of moderately dense  optically thick molecular clouds (see
Garc\'{\i}a-Burillo et al. 1993). The only exception to this is in the vicinity of the AGN
(R$<$50-100\,pc), where the measured ratio is close to 1. The highest values ($\sim$1.1) are
observed toward the lopsided structure in the CND which coincides with the peak of emission
detected in HCN (Helfer \& Blitz 1997). Ratios are close to 0.8--0.9 in the strongest clumps along
the spiral ridge of Arm I.

\section{Star formation in NGC\,4826 \label{starform}}

\subsection{Distribution of star formation \label{dist}}

The distribution of star formation in NGC\,4826's disk is 
strongly asymmetrical, as seen in the HST Pa$\alpha$ image (see Fig.~\ref{f9}). The asymmetries
follow with similar spatial scales the various $m$=1 instabilities identified in the CO maps. 
Interestingly, the overall star formation pattern shows a strong N/S asymmetry. While the
$^{12}$CO(1--0) emission on Arm II is associated with a strong ionized gas emission, to the southern
far
side in Arm I, where extinction is smaller, Pa$\alpha$ is significantly weaker. This is contrary to
what would be expected if the reported N/S asymmetry was due mainly to heavy extinction by the
conspicuous northern dust lane. Furthermore, the star formation pattern is also very asymmetrical
in the CND (Fig.~\ref{f9}).
We illustrate this in Fig.~\ref{f12}, which compares the Pa$\alpha$ morphology with the mostly
thermal radio continuum emission at 2cm in the nucleus of NGC\,4826 (from Turner \& Ho 1994). A
strong maximum is visible in both maps, which indicates that the star formation peaks neither on
the AGN nor on the lopsided instability seen in the $^{12}$CO(2-1) map, but rather 20\,pc
northward.

\subsection{Extinction in the ``Evil Eye" \label{ext}}

The conspicuous dust lane identified in the optical pictures of NGC\,4826 is expected
to significantly screen any emission to the N. With the assumption of intrinsic
axisymmetry for the stellar disk/bulge, Witt et al. (1994) determined the optical depth
in the ``Evil Eye". Following Witt et al. (1994), 
we have re-examined the issue of extinction by extracting surface brightness profiles
along the minor axis in F450W, F814W, and F160W. The relationship of this ``cut'' to the galaxy
morphology is shown in Fig.~\ref{galaxy}; the profiles, averaged over 3$\arcsec$ perpendicular to
the direction of the cut, are shown in Fig.~\ref{galaxy}.
The figure shows clearly the effects of extinction towards positive (northern)
radii, and confirms the assumption that the stellar populations in the disk
of NGC\,4826 are rather symmetric, as the discrepancy between the northern and
southern halves of the surface brightness profiles decreases substantially with increasing
wavelength. However, the ionized gas emission is clearly intrinsically
asymmetric since it is more prominent on the near side where there
is more extinction.

We have derived the extinction map using the mean colors of the disk,
as determined from the outer region of the cuts, toward the S,
where (broadband) extinction is low or zero (see Witt et al. 1994).
The $B-I$ mean color of the outer (S) region is 1.62, similar to,
although 0.16~mag bluer than, the equivalent color from de Jong (1996) for integrated colors
of similar morphological types. With the extinction coefficients given in Holtzman et al. (1995)
for the WFPC2 filters, we then derived the extinction from the $B-I$ map.
The mean extinction measured in the dust lane turns out to be not greater 
than $A_V\,=\,$1.5\,mag, in agreement with the findings of Walterbos et al. (1994)
and Witt et al. (1994). This confirms that the Pa$\alpha$ image of NGC\,4826 is virtually
extinction-free, and can be considered as a fair unbiased picture of how recent star formation
proceeds in
the central kpc of NGC\,4826.

\begin{figure}[!th]
   \centering
   \includegraphics[width=8.8cm]{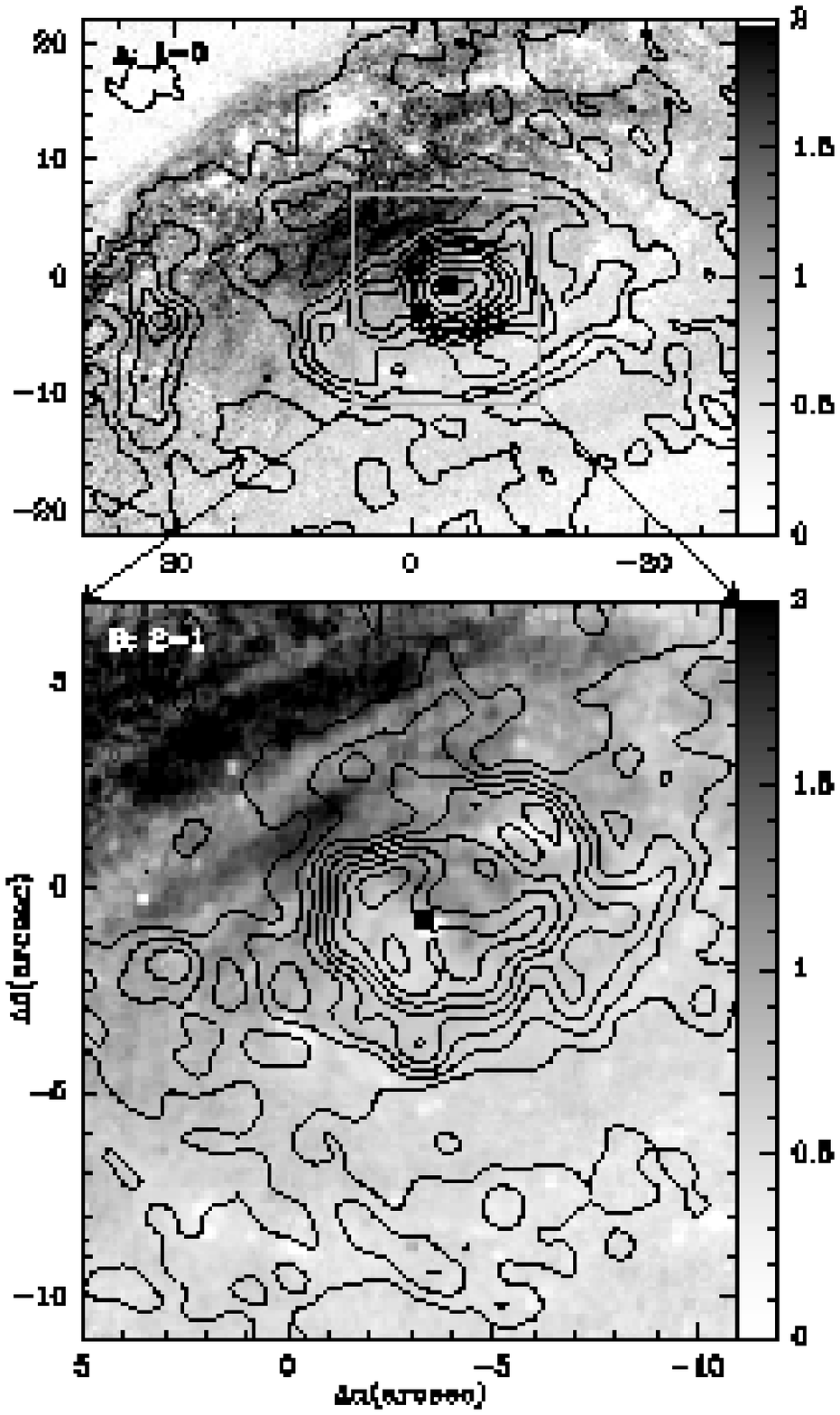}
      \caption{Same CO contours as in Fig.~\ref{f9} overlaid on extinction  A$_V$ derived
from the HST broad-band $B-I$ color map. Greyscale ranges from A$_V$=0 to 2.}
         \label{f11}
   \end{figure}

The average extinction in the nucleus of NGC\,4826 can also be estimated from the 
$^{12}$CO(1--0) integrated intensity map of Fig.~\ref{f3}. Assuming a CO-to-H$_2$
conversion factor $X$=N(H$_2$)/I$_{CO}$=2.2$\times$10$^{20}$~cm$^{-2}$~K$^{-1}$~km$^{-1}$~s
(Solomon \& Barrett 1991) and the widely used
gas-to-dust ratio of N(H$_2$)/A$_V\,=\,1.0\,10^{21}$~cm$^{-2}$\,mag$^{-1}$
(Bohlin et al. 1978), we obtain a mean $A_V\,\sim\,$10~mag
within the 42$\arcsec$ 1--0 primary beam. This value is a factor of 6 higher than
that derived from $B-I$ color. However, for several reasons the CO-based $A_V$ value 
likely overestimates the dust column density that can absorb efficiently the stellar
light. First, when this comparison is made, we implicitly assume a screen geometry for the H$_2$
slab with unit filling factor. Second, there is evidence that
the CO-to-$H_2$ column density ratio $X$ in central regions of galaxies may be lower 
than the standard Galactic value (see Garc\'{\i}a-Burillo et al. 2000 and references
therein). The extinction `discrepancy' is even more severe at small scales. Fig.~\ref{f11}
compares the extinction map, estimated from $B-I$, with the CO maps. 
Confirming the impression given in Fig.~\ref{galaxy}, it is evident
that molecular gas column density is not tightly correlated with the $B-I$-based
$A_V$ (see also Block et al. 1994 and Sakamoto et al. 1999).    
Moreover, the clumpy structure of the interstellar medium makes difficult any comparison
between $A_V$ values derived using data taken at very different spatial resolutions.

\subsection{The star formation rate \label{sfr}}

We have measured the Pa$\alpha$ flux within the entire NICMOS field-of-view for NGC\,4826, 
summing $>$\,3$\sigma$ above sky, and found F(Pa$\alpha$)\,=\,7.4$\,\times\,10^{-13}$ erg/s/cm$^2$.
This gives a luminosity L(Pa$\alpha$)\,=\,1.5$\,\times\,10^{39}$ erg/s,
or a massive star formation rate (SFR) of 0.14\,M$_\odot$/yr, after correcting
for a mean extinction $A_V$ of 1.5 mag (this work).
This is slightly lower than, although comparable to, the SFR of 0.22 \,M$_\odot$/yr
estimated by Braun et al. (1994) from an H$\alpha$ image with a larger field-of-view.
While this indicates rather vigorous star formation activity, it is much lower
than in true nuclear starburst galaxies such as M\,82, which have SFRs more than
10-50 times higher (O'Connell et al. 1995).

The number of ionizing photons can also be measured from the ionized gas emission,
and from this we can derive the number of equivalent O stars necessary to ionize
the gas. Using the coefficients given in Osterbrock (1989), we find that
the total Pa$\alpha$ luminosity (not corrected for extinction)
corresponds to 9.3\,$\times\,10^{51}$ ionizing photons.
Taking an O7V star as representative, and assuming the number of
Lyman continuum photons from such a star to be 1$\,\times\,10^{49}$
(Leitherer 1990), this corresponds to roughly 930 O7V stars.
Our estimates of ionizing photons are consistent with
those of Pierini et al. (2002), if we consider the total area of the Pa$\alpha$
image compared to that of their spectroscopic slit.
Because extinction may be an issue here, it is important to compare the number of
ionizing photons derived from Pa$\alpha$ to that derived by Turner \& Ho (1994) from the radio 
continuum. They find a factor of 3 fewer ionizing photons, but 20 times more massive
stars, differences which almost certainly are due to the different normalizations. As noted by
Turner \& Ho (1994), the H{\sc II} regions in NGC\,4826 are one order of magnitude more energetic
than the largest star-forming complexes in the Milky Way.
In any case, the rough agreement between the estimates obtained from Pa$\alpha$ and from radio 
continuum maps is an independent confirmation that Pa$\alpha$ suffers from virtually no extinction
(see also discussion in Sect. \ref{ext}). 

   \begin{figure}
   \centering

   \includegraphics[angle=0,width=8.5cm]{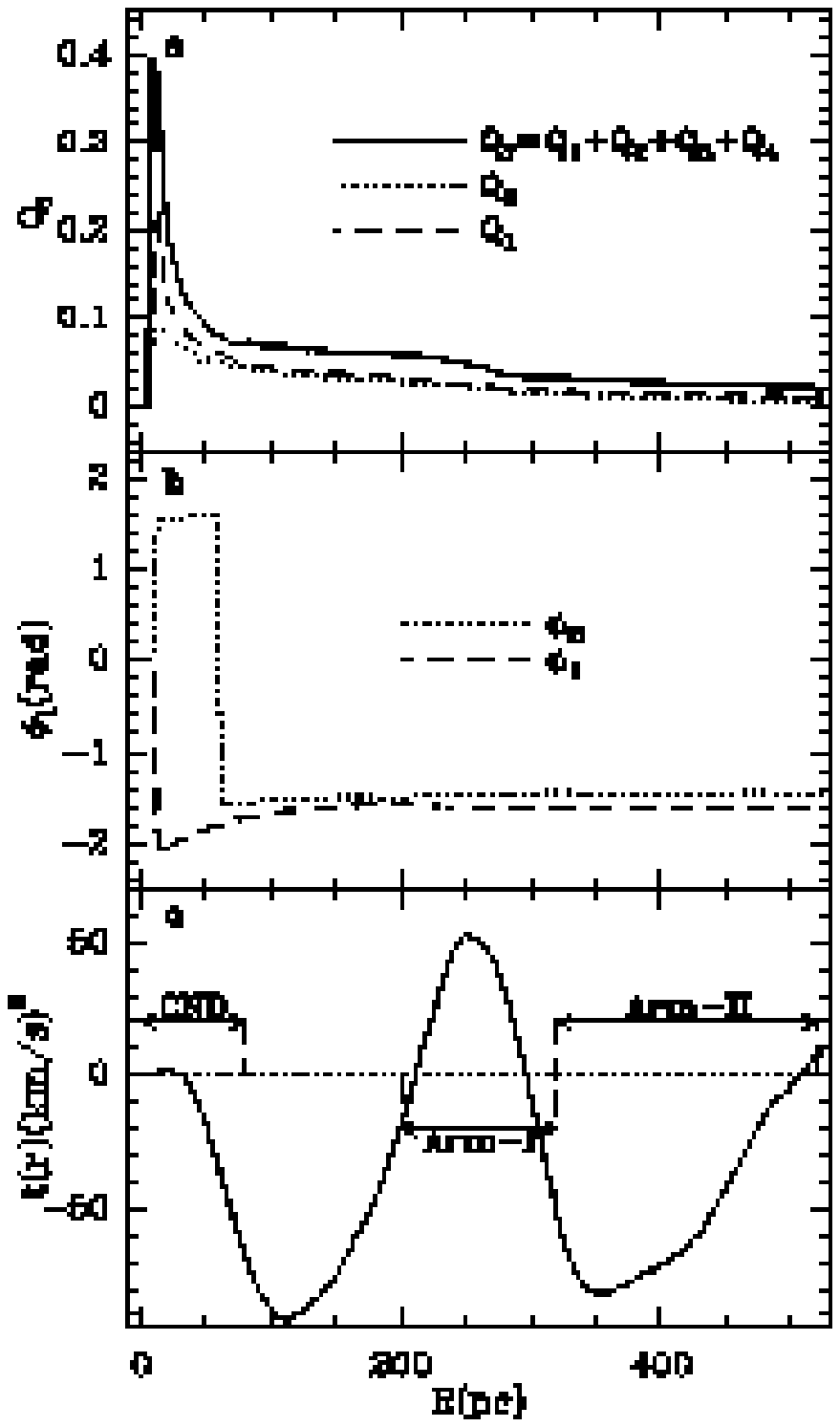}

      \caption{{\bf a(top)}Strength of the i-Fourier components of the stellar potential of 
NGC\,4826 derived from the HST-NICMOS image of the galaxy nucleus. We represent the main
components--Q$_{1}$ and Q$_{2}$--and the sum Q$_{T}$. {\bf b(middle)} The phases $\Phi_{i}$ of the
potential components are plotted for $i$=1,2.{\bf c(bottom)} Radial variation of the average 
torque--$t(R)$--per unit mass. We indicate the radial extent of CND, Arm I and Arm II.}
         \label{f14}
   \end{figure}
\subsection{Star formation and compact star clusters in NGC\,4826 \label{scc}}
  
The comparison of the Pa$\alpha$ morphology with that of the F450W
and the $B-I$ color map reveals an interesting feature.
The very blue compact source that emerges in the color cut at 17$\arcsec$
lies within a ridge of H{\sc II} regions or compact star clusters, visible
in Fig.~\ref{galaxy} just inside the edge of the ``Evil Eye''.
Surprisingly, this ridge of star formation does not coincide with the 
ionized gas emission ridge seen in Pa$\alpha$ (Fig.~\ref{f9}).
The Pa$\alpha$ ridge is positioned roughly 4$\arcsec$ closer (80~pc) to the nucleus 
than the arc of clusters, while the (very blue) 17$\arcsec$ feature seems to contain 
little if any ionized gas.
This result is consistent with the radial variations of ionized gas emission
along the slit of Pierini et al. (2002), who found the peak of the ionizing
photon flux at $\sim\,14-15\arcsec$; at 17$\arcsec$ the ionizing flux
is approximately 30\% of its peak value. 
These authors also found the hardness of the ionizing ratio to vary radially,
with an abrupt softening at 17\,$\arcsec$, roughly the same location where
we find the blue compact source in the color cuts.

To better understand the pattern of star formation in NGC\,4826,
we have measured F450W and F814W magnitudes and (FWHM) diameters
of compact sources in the ridge just inside 
the edge of the dust lane (see extra-nuclear cross in Fig.~\ref{galaxy}).
All of the 25 sources measured appear to be star clusters, since their
absolute F814W\,$\sim\,I$ magnitudes range from $-7.4$ to $-8.9$
(without correction for extinction) and their FWHMs from 4 to 29\,pc.
If we convert to $V$ assuming $B-V\,\sim\,0.7$ and correct for 1.5 mag of
visual extinction, we find that the absolute $V$ magnitudes ($-8.5$ to $-9.9$)
are consistent with those of Super Star Clusters (SSCs) if they are old enough (in which
case the correction to the fiducial age of 10\,Myr would be $>$ 1\,mag).

If we infer the ages of these clusters from their $B-I$ color,
which ranges from 0.65 to 2, we would derive a minimum age of 40\,Myr
(Leitherer et al. 1999) in the absence of reddening.
This would imply a minimum luminosity correction of $\sim\,1.5$\,mag 
(Whitmore et al. 1997), which would bring the luminosity of most of these
compact star clusters into the SSC range of M$_V\,<\,-10.5$ at a fiducial
age of 10\,Myr (Billet et al. 2002). These clusters are not extremely compact,
although this also  could be related to an older age. In any case, the clusters must be older than 
\,10\,Myr, given the conspicuous absence of ionized gas emission in their 
immediate vicinity. The formation scenario for compact star clusters or SSCs,
although not completely understood, appears to involve mergers (Bekki \& Couch 2001) and/or
high-pressure environments (e.g., Billet et al. 2002).

The ionized and the molecular gas profiles along the minor axis of NGC\,4826 share a
common feature: the disks are abruptly truncated beyond a ridge of old ($>$10\,Myr) compact star
clusters. This pattern suggests a parallel pattern in the age of the star formation in 
the disk. Apparently stars are still vigorously forming elsewehre inside R$\sim$700\,pc where there
is a significant molecular gas reservoir.

\section{Counter-rotation and $m$=1 modes \label{counter}}

NGC\,4826 is a prototype among the large variety of counter-rotating disk galaxies. 
Counter-rotation can be purely stellar, involving stars versus gas, or characterized by the
co-existence of two decoupled gas disks as seen in NGC\,4826 (see compilations by Bertola \&
Corsini 1999, and Bettoni et al. 2001). The existence of counter-rotating components in disks
suggests the active role that interactions, minor mergers and mass accretion processes may have in
driving the evolution of galaxies. In hierarchical merging scenarios which explain how galaxy disks
are built up, counter-rotating galaxies may naturally result from merger and/or 
accretion events. Partly due to the small number of observations, the prevalence of 
decoupled components in disk galaxies may have been underestimated thus far. 
Nevertheless, it is important to investigate which are the main gravitational
instabilities induced by counter-rotation, and to analyze their influence on the removal of angular
momentum from the gas. Analytical studies by Lovelace et al. (1997) indicate 
that the main dynamical instabilities linked with two-stream flows in galactic disks
are one-arm spirals, i.e., $m$=1 modes.

Garc\'{\i}a-Burillo et al. (1998, 2000) mapped at high-resolution the massive 
counter-rotating molecular disks of the spirals NGC\,3593 and NGC\,3626, searching
for the signatures of counter-rotating instabilities. The maps of 
both galaxies show compact molecular gas disks with mixtures of 
$m$=1 and $m$=2 perturbations. With the help of self-consistent numerical simulations adapted for 
NGC\,3593, Garc\'{\i}a-Burillo et al (2000) studied the development of disk
instabilities. In their study, they found that counter-rotation drives rapidly
evolving perturbations. The disk first develops stationary
($\Omega_p\sim$0~km\,s$^{-1}$~kpc$^{-1}$) waves, leading with respect to the gas flow, followed
later
by a mixture of $m$=1 and $m$=2 slow (trailing) waves. The nature of the instabilities is seen to
depend critically on the assumed halo/disk mass ratio (see also Comins et al. 1997). 
Similarly to the aforementioned cases, NGC\,4826 has a very compact molecular disk 
with signatures of $m$=1 perturbations at various scales.
Rix et al. (1995) analyzed the stellar velocity distributions and estimated that a
sizable fraction (10$\%$--30$\%$) of the stars in 
NGC\,4826's  inner disk (radii $<$50$\arcsec$) could be in counter-rotation. This
decoupled component could drive the two-stream flow instabilities we 
actually observe in the molecular disk. 
However, the reported fraction of counter-rotating stars is still compatible with 
the high-velocity dispersion of this kinematically hot region, and thus cannot be
taken as firm evidence for counter-rotation in the stellar component.

In summary, the only {\it true} counter-rotating component in NGC\,4826 may be the
outer HI disk. The instabilities observed in the inner molecular disk are therefore not linked with
counter-rotation: $m$=1 perturbations are detected in a region where two-stream 
flow is not detected.  Additional evidence for this comes from the 
analysis of streaming motions made in the frame of density wave models (see Sect. \ref{modes}). The
CND lopsided instability and Arm I behave as fast modes, i.e., if the inner $m$=1 perturbations
are wave modes they seem to have developed outside corotation. This contrasts with the low pattern
speeds of $m$=1 modes which are known to be generated by the two-stream flow instability (Lovelace
et al. 1997; Comins et al. 1997; Garc\'{\i}a-Burillo et al. 2000).

\section{Gravitational instabilities and AGN fueling \label{grav}}

\subsection{Gravitational torques in NGC\,4826}

The objective of this section is to study if gravitational torques, derived from a fair
representation of the stellar potential in the inner disk of NGC\, 4826, can account for the 
gas kinematics derived from CO. Furthermore, we examine the efficiency of gravitational
torques exerted on the gas in the context of AGN fueling.

NIR maps can be used to derive the distribution of old stars as they are less affected by dust
extinction or by stellar population biases. Here we have used a high-resolution near-infrared image
of NGC\,4826 taken with HST-NICMOS to derive the stellar potential within the inner 1\,kpc of the
disk. The HST field-of-view on NGC\,4826 is 1024\,pc and is sampled with a grid 
of 256x256 pixels of 0.203$\arcsec$ spatial resolution.
The image of the galaxy was completed along the 3rd vertical dimension by assuming
an isothermal plane with a scale height constant with radius; this scale height
is equal to 1/12th of the radial scale-length, i.e.,$\sim$160~pc.
The gravitational potential is derived by a Fourier transform
method on the grid from the NIR image. We also assumed a constant
mass-to-light (M/L) ratio; the value of M/L is obtained by
fitting the observed rotation curve.

The potential is decomposed as
\begin{equation}
\Phi(R,\theta) = \Phi_0(R) + \sum_m \Phi_m(R) \cos (m \theta - \phi_m)
\end{equation}
we define the strength of the $m$-Fourier component,
$Q_m(R) = m \Phi_m / R | F_0(R) |$ and its global strength
over the disk as $\max_R Q_m(R)$ (e. g. Combes \& Sanders 1981). 

The strength of the total potential or maximal torque over the whole disk is defined
by
\begin{equation}
Q_T(R) = {F_T^{max}(R) \over F_0(R)} = 
{{{1\over R}\bigl{(}{\partial \Phi(R,\theta)\over \partial\theta}\bigr{)}_
{max}} \over {d\Phi_0(R)\over dR}} 
\end{equation}
\noindent
where $F_T^{max}(R)$ 
represents the maximum amplitude of the tangential force
and $F_{0}(R)$ is the mean axisymmetric radial force, inferred
from the $m$=0 component of the gravitational potential. 

The strengths of the main $m$ components for R$<$520~pc are plotted in Fig.~\ref{f14},
together with the radial-variation of the corresponding phases, $\phi_m$.
It is interesting to notice that $m$=1 components are stronger than $m$=2 components
in the stellar potential inside this region, especially for R$<$50~pc. This indicates that
there are also asymmetrical perturbations in the stellar disk. 

 Once the potential is obtained all over the grid, we deduce the
forces ($F_x$ and $F_y$) by derivation in each pixel, and can compute the average
torque exerted on the gas.  The surface  density $\Sigma$ of the gas is assumed
proportional  to the CO emission, either $^{12}$CO(1-0) or $^{12}$CO(2-1) (both
maps are used). We resample the CO maps at the same pixel size of
0.203", and deproject them with the same geometrical angles as before.
On each pixel, the torque $T$ is computed by:

$$
T = \Sigma (x.F_y -y.F_x)
$$

We then compute the average over each radius of the torque per unit mass,
i.e.:

$$
t(R) = \frac{\int_R \Sigma (x.F_y -y.F_x)}{\int_R \Sigma }
$$

The mean torques calculated using the $^{12}$CO(1-0) map are plotted in Fig.~\ref{f14}, in units of
$(km/s)^2$. The results based on the $^{12}$CO(1-0) and $^{12}$CO(2-1) maps are consistent. Most
remarkably, oscillations between negative and positive values for $t(R)$ are roughly consistent
with the diagnostic based on the change of sign of streaming motions for Arm I (related to a 
fast trailing wave, hence causing angular momentum gain) and Arm II (related to a 
slow trailing wave, hence causing angular momentum loss) (see Sect. \ref{modes}). The corresponding
stellar torques in the CND are marginally positive very close to the AGN (R$<$50~pc), suggesting
that stellar perturbations contribute little to AGN feeding. Again,
this fits 'qualitatively' the  streaming motions measured in CO, which would rather suggest 
outflow to some degree. The stellar torques change to negative values in the outer boundary of the
CND; this roughly agrees also with the CO-based diagnostic.

While the radial variation of the stellar torques seem to account qualitatively for the changing
signature of streaming motions, the maximum value of the mean
torque is exceedingly small, however: $\sim$50$(km/s)^2$. This implies that 
the typical time-scale for the gas to lose its angular momentum for R$\sim$200\,pc is 
of about 1 Gyr. This is much longer than the gas dynamical time-scale at this radius
(of the order of 2 Myr). Gravitational torques exerted on the gas due to the 
stellar potential are 'globally' quite weak. Weighted by the density, the total torque inside the
evaluated region of the disk (R$\sim$520~pc) is only marginally negative: the combination of all
stellar perturbations seems to make the gas lose its angular momentum rather inefficiently.

We can ask whether our conclusions might be partly biased by the
existence of residual extinction in the $H$-band image originally used to
derive the gravitational potential.  However, several arguments can be
advanced suggesting that this bias is negligible.  Although we estimate that
extinction in the $H$ band can locally reach $A_{H}\sim 0.20-0.25$ magnitudes
in the northern dust lane arc (at $R \sim 250-300\,{\rm pc}$), this value is
comfortingly low elsewhere.  In particular, we do not expect that the stellar
gravitational potential is severely biased inside the CND, i.e., at the scales
which are critical for evaluating the influence of gravitational torques on
the AGN feeding. The derived values for $t(R)$ inside the CND (at $R <
80\,{\rm pc}$ where $A_{H}< 0.02-0.03$), as well as in Arm I (developing on
the southern side, where $A_{H} < 0.01$), are mostly weighted by the
contribution of regions that are virtually extinction-free (see
Fig.~\ref{f11}).

To estimate quantitatively the global influence of extinction on the derived
torques, we have corrected the NIR map by a factor exp($\tau_H$), where
$\tau_H=0.14 A_V$ and $A_V$ corresponds to the extinction map of
Fig.~\ref{f11}. Confirming our expectations, the Fourier analysis of the
gravitational potential thus derived is only slightly modified, and the
average torques on the gas are practically unchanged.  Although a detailed
evaluation of residual extinction on the $H$-band image would require the use
of several high-resolution optical and infrared images in order to resolve the
age-extinction degeneracy (to be addressed in a forthcoming publication), we
can conclude that the value derived here for $t(R)$ in the
center of NGC\,4826 is a robust estimate.

   \begin{figure}[!t]
   \centering
   \includegraphics[width=8.8cm]{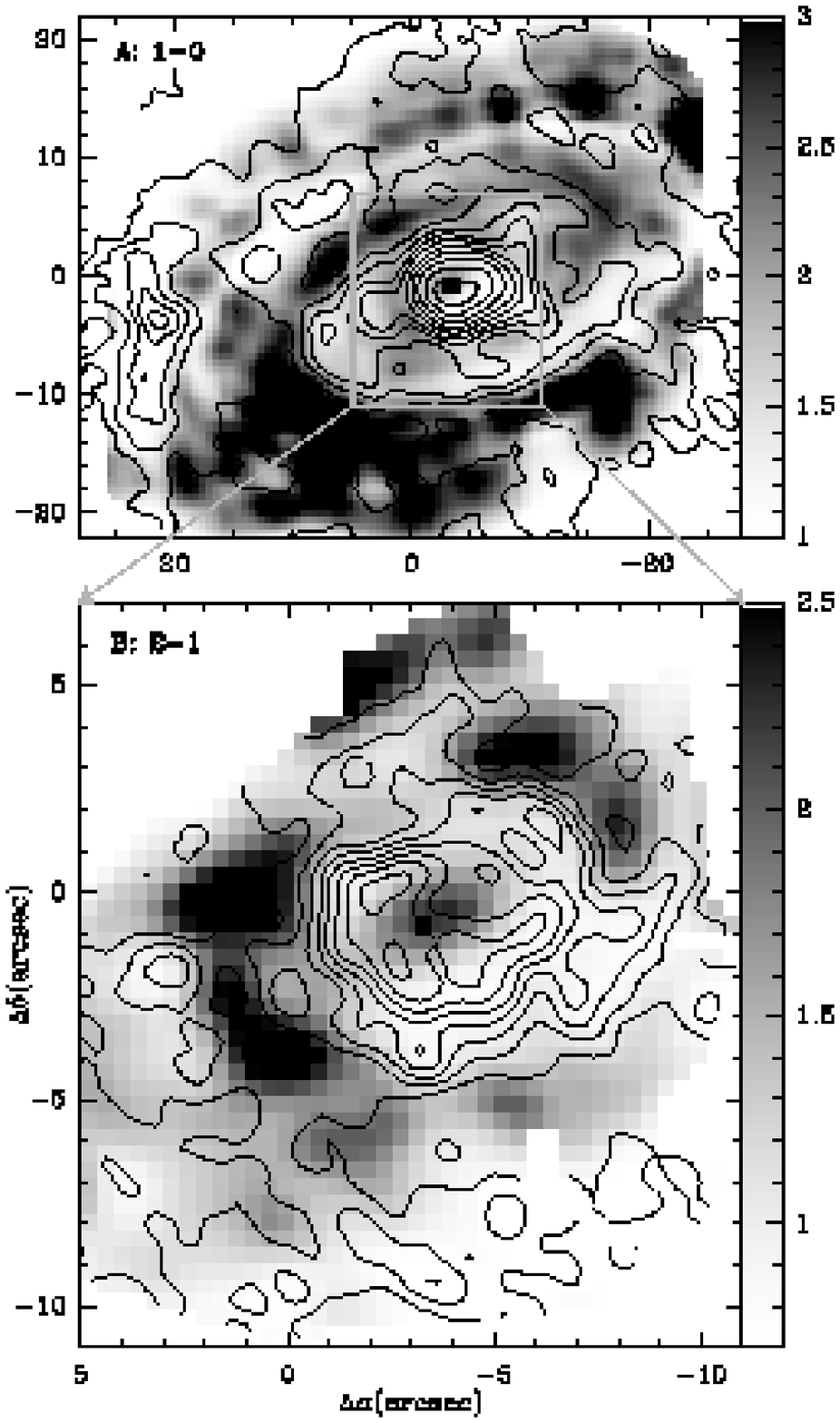}

      \caption{Estimation of the order of magnitude of the Toomre $Q$ parameter (grey
scale ranges from  1--3 in the top panel and from 0.7--2.5 in the bottom panel) overlaid on the 
integrated intensity maps obtained for the 1--0 (top) and 2--1(bottom) lines of
$^{12}$CO in the disk of NGC\,4826 (same contours as Fig.~\ref{f5}).}

         \label{f13}
   \end{figure}

   \begin{figure}
   \centering
   \includegraphics[width=8.5cm]{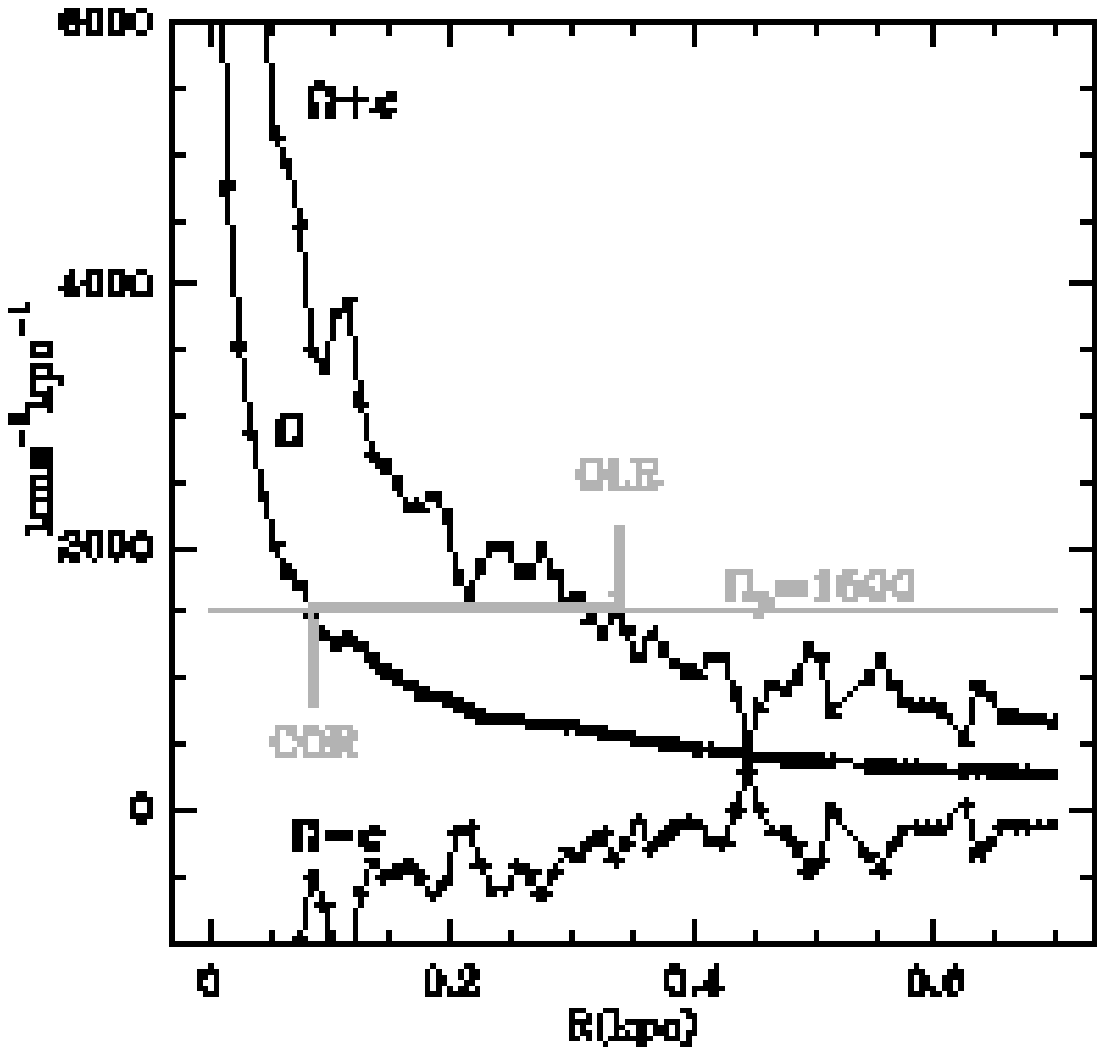}
      \caption{The principal epicyclic frequencies $\Omega$, 
$\Omega$-$\kappa$ and $\Omega$+$\kappa$ in the inner disk of NGC\,4826. The pattern speed 
$\Omega_p$=1500\,km~s$^{-1}$kpc$^{-1}$ has been purposely chosen to locate the inner
m=1 mode of Arm I between its corotation and OLR.}
         \label{f15}
   \end{figure}

\subsection{Gas self-gravity}

Stellar perturbations are weak in the inner disk of NGC\,4826 (except in the very center of the
galaxy: R$\leq$50~pc). Furthermore, we have found indications that they are probably inefficient to
drive AGN fueling at R$\leq$520~pc. Alternatively, self-gravitating gas instabilities may have a
potential role in AGN feeding. When self-gravity of stars and/or gas is strong, galaxy disks are
prone to gravitational instabilities. The local stability criterion, first established by Toomre
(1964), ensures stability against axisymmetric perturbations 
in a stellar/gaseous galaxy disk when the column density of stars/gas ($N_{*,g}$) is
smaller than a critical value ($N_{*,g}^{crit}$=$\kappa\sigma_v$/$\pi G$), i.e., when the
so-called Toomre parameter $Q_{*,g}$=$N_{*,g}^{crit}$/$N_{*,g}>1$. Here, $\kappa$ is the epicyclic
frequency, $\sigma_v$ is the velocity dispersion and G is the gravitational constant. Stability
against non-axisymmetric perturbations, such as spirals and bars, is guaranteed when $Q$ is
significantly larger than 2.

The stability criterion for a realistic galaxy disk needs to take into account the
strong coupling between the stellar and gaseous components through self-regulation and
feedback (Combes 2001). Nevertheless, as a first step, it is useful to evaluate $Q$ {\it separately}
for the gas disk. In the particular case of AGN host disks, previous extinction maps derived from
HST color images seem to indicate that non self-gravitating gas instabilities
($Q\sim$10-100) may be ubiquitous in the inner 100-200\,pc of Seyfert galaxies (Martini \& Pogge
1999) and potential drivers of AGN fueling through energy dissipation in shocks and turbulent
motions. This scenario can be fully tested by exploiting the information provided by $^{12}$CO
maps. Under certain assumptions, the $^{12}$CO intensity gives a fair estimate of
$N_{g}$. Moreover, the $^{12}$CO kinematics can be used to derive $\Omega$, $\kappa$, and $\sigma_v$
in the disk. Hence, the derivation of $Q$ maps follows straightforwardly.

Fig.~\ref{f13} shows the variation of $Q$ in the molecular disk of NGC\,4826 as derived
from $^{12}$CO(1--0). Fig.~\ref{f13} displays also a zoomed-in view of the $Q$ map for the inner
disk obtained from $^{12}$CO(2--1). Gas column densities $N_{g}$ are derived from 1--0 data as
explained in Sect. \ref{gmass}. We have used the 2--1/1--0 line ratio maps discussed in Sect.
\ref{ratios} to calculate $N_{g}$ from $^{12}$CO(2--1). $\Omega$ and $\kappa$ are derived from
v$_{rot}$ using standard definitions. Finally, $\sigma_v$ has been estimated from CO second moment
maps. The contribution to $\sigma_v$ of the rotation curve gradient within the lobe(s) is found
to be significant for $r<2\arcsec$, but negligible elsewhere. The velocity dispersion is fairly
constant in NGC\,4826's disk and close to 10~km~s$^{-1}$ except for $r<2\arcsec$, where $\sigma_v$
doubles its value on average. In the following analysis we take for simplicity
$\sigma_v$=10~km~s$^{-1}$.

With these assumptions, Fig.~\ref{f13} indicates that Arms I and II are self-gravitating
perturbations: the two one-arm spirals are characterized by low $Q$ values, ranging from 1 to 2.
There are also indications for the CND lopsided instability to be self-gravitating ($Q<$2--3),
unless an unrealistic boost in velocity dispersion (not observed) or a much lower CO-to-H$_2$
conversion factor is invoked for $r<2\arcsec$. NGC\,4826 seems to be a good counterexample to the
scenario suggested by Martini \& Pogge 1999: the $m$=1 instabilities identified in NGC\,4826 have
low Q values. This suggests that self-gravity of the gas cannot be neglected and probably plays an
essential role in the maintenance of the $m$=1-type perturbations described in NGC\,4826.

\section{Summary and conclusions} 

We summarize the main results obtained here as follows:

\begin{itemize}

\item

High-resolution $^{12}$CO observations of NGC\,4826 show that the bulk of its
molecular gas lies in a highly structured disk of
M$_{gas}\sim$3.1$\times$10$^{8}$M$_{\sun}$ which
ends abruptly at R$\sim$700~pc. 
The gas disk shows a preponderance of asymmetric perturbations ($m$=1 instabilities). 
There is a lopsided nuclear disk of 40\,pc radius and two one-arm trailing spirals, 
which develop at different radii in the disk (Arm I:~200--350~pc, Arm 
II:~350--700~pc). 

\item 

 The star formation pattern in the disk of NGC\,4826 is strongly asymmetrical. The
scales of the observed asymmetries resemble those of the various $m$=1 perturbations 
revealed in the disk of molecular gas. Furthermore, the star-forming ionized/molecular gas disk is 
truncated beyond a ridge of compact star clusters identified at
R$\sim$700~pc. This suggests an evolutionary trend for star formation in the disk of NGC\,4826.
Apparently massive star formation has ceased beyond R$\sim$700~pc, but is still vigorously occurring
inside this radius, fed by the significant molecular gas reservoir.

\item

Gas kinematics reveal streaming motions related to the $m$=1 perturbations. A first analysis of
these perturbations suggests that the inner $m$=1 instabilities may be fast trailing waves which
have developed between corotation and an Outer Lindblad Resonance (OLR). This would imply that the
AGN is probably not being generously fueled in the current epoch. An estimate of the radial
variation of the mean gravitational torques due to the stellar potential confirms independently that 
stellar perturbations are inefficient to drive AGN fueling.

\item 

Arms I and II are both self-gravitating perturbations characterized by low values of
the Toomre $Q$ parameter ($Q$=1--2). There are also indications that the CND lopsided
instability is self-gravitating ($Q<$2--3).

\end{itemize}

The two-stream flow instabilities, expected in counter-rotators, cannot explain the ubiquity of
$m$=1 perturbations in the inner disk of the {\it Evil Eye}.  Mechanisms other than
counter-rotation have been suggested to trigger $m$=1 instabilities, such as interactions with
companions (Weinberg 1994;  Lovelace et al. 1999), central potentials dominated by massive black
holes (Miller \& Smith 1992; Taga \& Iye 1998; Bacon et al. 2001) and response to an asymmetric
halo (Jog 1997). Although pure $m$=1 modes are uncommon in galaxy disks, kinematic lopsidedness has
been reported in 50$\%$ of spirals (Richter \& Sancisi 1994). The self-consistent
numerical simulations made by Junqueira \& Combes (1996) showed that above a threshold  
central gas concentration, a typical galaxy disk (including stars and gas)
is prone to develop one-arm (trailing) spiral perturbations. The strongest 
$m$=1 modes in their models appear between their corotation and their OLRs; in other
words, these perturbations are fast. Fast $m$=1 trailing modes are decoupled from outer disk
perturbations which only have 1/10 the speed of the inner waves. 
According to Junqueira \& Combes (1996), the fast $m$=1 modes might reflect the
action of the modal amplification mechanism described by Shu et al. (1990). 
The modes are excited by a slight off-centering of the stars and gas
in the galaxy nucleus and develop mainly in the gas disk
between corotation and OLR. In summary, the inner $m$=1 perturbations identified in the gas
disk of NGC\,4826 could be tentatively explained by this scenario if the pattern speed of the
perturbation is tuned to high enough values.
As shown in Fig.~\ref{f15}, a value of $\Omega_p\geq$1500~km~s$^{-1}$~kpc$^{-1}$ would push
corotation well inside the disk, accounting for the spatial extent of Arm I in
NGC\,4826.

\subsection{A possible scenario for the evolution of the inner kiloparsec of
NGC\,4826}

While it appears that counter-rotation is not driving $m=1$ modes in the inner
1.5~kpc disk of NGC\,4826 at present, a past gas accretion episode must be invoked to explain the 
decoupling of HI in the outer disk.   
Furthermore, the formation of compact/truncated star forming disks in
counter-rotators, such as the one we see in NGC\,4826, could represent the final stage of a process
involving large-scale collisions between the accreted gas and the primary gas of the accretor
(Rubin 1994, Thakar et al 1997, Garc\'{\i}a-Burillo et al 2000). In general, the two components may
have opposite angular momenta, implying substantial dissipation when they mix and, 
eventually, transformation from atomic to molecular phase in shocks. Large amounts of
gas may fall towards the nucleus and form a circumnuclear gas disk, whose final size will
depend on the ratio of initial angular momenta of the two components. In close agreement
with this picture, the kinematics of ionized gas reveal an orderly infall of gas from 
R=2~kpc to 800~pc in NGC\,4826 (Rubin 1994). 
Moreover, the measured [NII]/H$\alpha$ ratio is close to 1 in this 
transition region, strongly suggesting shock excitation (Rubin 1994).
In the course of this process, massive star formation is triggered along the
``ridge'' and in the nuclear disk. The 
time-scale for gas infall may be short and similar to the dynamical time, i.e., 
$\sim$10$^{7-8}$yr. This estimate is consistent with the age of the stellar clusters inferred
from their $B-I$ colors. 

The stellar velocities measured in the nucleus of NGC\,4826 also imply a recent large
mass infall episode. Kormendy (1993) has discussed NGC\,4826 as an example of an
early-type spiral with an anomalously low stellar velocity dispersion for its bulge luminosity 
($\sigma_v$=90$\pm$5~km~s$^{-1}$, later confirmed by Rix et al 1995). The expected
value of $\sigma_v$ for NGC\,4826 predicted by the Faber-Jackson relation would be
$\sim$160~km~s$^{-1}$. As argued by Kormendy (1993), this indicates that the central brightness is
dominated by a cold disky component.  The formation of this cold central disk, in contrast to
standard hot bulges, might be related to secular evolutionary processes involving large mass
accretion.

Once the gas has settled in the nuclear disk, normal secular evolution can proceed.
The onset of $m$=1 instabilities of the type described in NGC\,4826 may be a
consequence of secular evolution in disks with large gas masses. Detailed numerical simulations to
be presented in a forthcoming paper (Garc\'{\i}a-Burillo et al. 2003 in prep) will study the onset
and evolution of asymmetric modes for a case similar to 
NGC\,4826, where the role of self-gravity of the gas may be essential.

\begin{acknowledgements}
         We acknowledge the IRAM staff from the Plateau de Bure and from 
         Grenoble for carrying out the observations and help provided during the 
	 data reduction. This paper has been partially funded by the Spanish MCyT 
         under projects DGES/AYA2000-927, ESP2001-4519-PE 
         and ESP2002-01693, and European FEDER funds.

\end{acknowledgements}

\end{document}